%% file: ultra_strong11.tex
\documentclass[aps,prb,preprint,groupedaddress,amsmath,amssymb, showpacs]{revtex4-1}
\usepackage{epsfig,psfrag}
\usepackage{dcolumn}
\usepackage{bm}
\usepackage{epsfig} 
\usepackage{graphicx}
\usepackage{epstopdf}
\usepackage{xcolor}
\usepackage{array}
\usepackage{import}
\usepackage{siunitx}
\usepackage{transparent}
\begin{document}
\title{Strong and ultra-strong coupling with free space radiation}
\author{S. Huppert}
\author{A. Vasanelli}
\email{Angela.Vasanelli@univ-paris-diderot.fr}
\author{G. Pegolotti}
\author{Y. Todorov}
\author{C. Sirtori}

\affiliation{Universit\'e Paris Diderot, Sorbonne Paris Cit\'e, Laboratoire Mat\'eriaux et Ph\'enom\`enes Quantiques, UMR 7162, 75013 Paris, France} 




\begin{abstract}
Strong and ultra-strong light-matter coupling are remarkable phenomena of quantum electrodynamics occurring when the interaction between a matter excitation and the electromagnetic field cannot be described by usual perturbation 
theory. This is generally achieved by coupling an excitation with large oscillator strength to the confined electromagnetic mode of an optical microcavity. In this work we demonstrate that strong/ultra-strong coupling can also 
take place in the absence of optical confinement. We have studied the non-perturbative spontaneous emission of collective excitations in a dense two-dimensional electron gas that supperradiantly decays into free space. By 
using a quantum model based on the input-output formalism, we have derived the linear optical properties of the coupled system and demonstrated that its eigenstates are mixed light-matter particles, like in any system displaying 
strong or ultra-strong light-matter interaction. Moreover, we have shown that in the ultra-strong coupling regime, i.e. when the radiative broadening is comparable to the matter excitation energy, the commonly used rotating-wave 
and Markov approximations yield unphysical results. Finally, the input-output formalism has allowed us to prove that Kirchhoff's law, describing thermal emission properties, applies to our system in all the light-matter coupling regimes considered in this work.  
\end{abstract}


\maketitle

\section{Introduction}
\begin{figure}
\centering  
\includegraphics[scale=0.7]{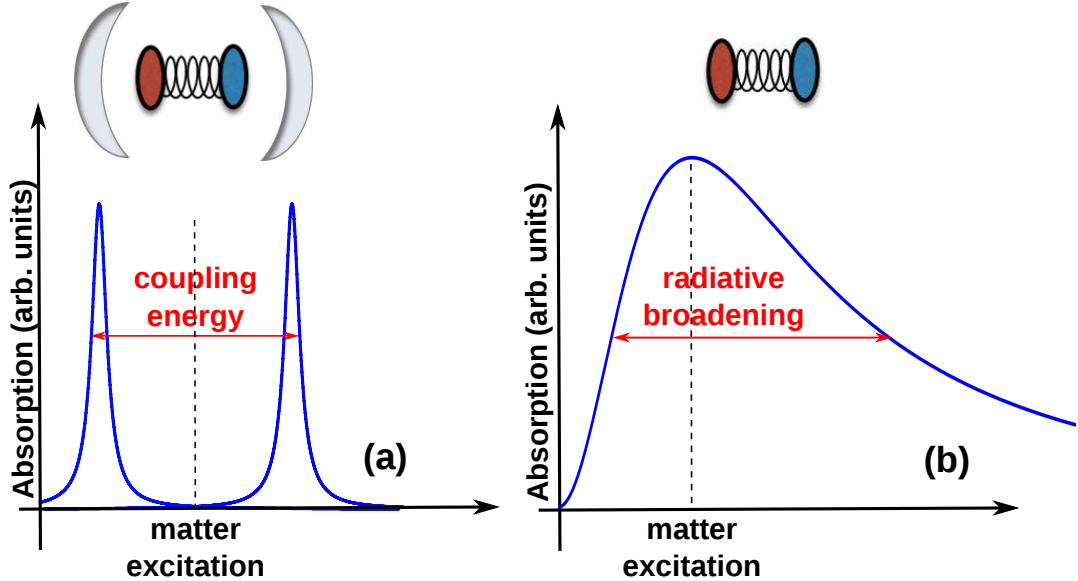}
\caption{Sketch of two systems in the ultra-strong coupling regime. In panel (a) a matter excitation is coupled with a resonant microcavity mode, giving rise to two new eigenmodes, the microcavity polaritons. In the ultra-strong 
coupling regime the energy separation between the polaritons, proportional to the coupling energy, is comparable with the matter excitation energy. Panel (b) sketches a matter excitation coupled with free space radiation, 
with a radiative broadening comparable to the matter excitation energy.}
\label{fig_introduction}
\end{figure}

Strong and ultra-strong light -- matter coupling regimes refer to systems in which light-matter interaction cannot be described within the framework of a perturbative theory.  
Most commonly, in condensed matter systems, these regimes are reached by inserting a material excitation at energy $E_{matter}$ into a resonant microcavity. 
Indeed, as the coupling energy $E_R$ is inversely proportional to the square root of the cavity 
volume, light -- matter interaction can be strongly enhanced by a tight photonic confinement and $E_R$ can become greater than the intrinsic material and photonic linewidths. This situation is schematized in fig.~\ref{fig_introduction}a: the non-perturbative nature of the coupling between the matter excitation and 
the resonant photon mode manifests itself through the appearance of two new mixed light -- matter states, the microcavity polaritons~\cite{WeisbuchPRL1992polariton}, whose energy separation is related to the coupling energy $E_R$. 
In the usual strong coupling regime ($E_R \ll E_{matter}$), the energy of the two polariton states 
at resonance varies linearly with the coupling energy. The ultra-strong coupling regime is obtained when the coupling energy is of the same order of magnitude as that of the matter excitation 
($E_R \approx E_{matter}$)~\cite{ciutiPRB2005ultrastrong}. In that case, the routinely invoked rotating-wave approximation (RWA) is no longer applicable and the light-matter interaction Hamiltonian has to include anti-resonant terms.
Due to these terms, the polariton branches are not linear anymore as a function of the coupling energy\cite{ciutiPRB2005ultrastrong}. The observation of such a non-linearity has been considered as the signature of the ultra-strong 
coupling regime in several material systems: superconducting 
circuits~\cite{niemczyk, ballester}, cyclotron transitions~\cite{scalari_science} and cyclotron plasma~\cite{muravev}, Frenkel molecular excitons~\cite{kena-cohen, gambino}, dye molecules~\cite{schwartz}, intersubband 
transitions in quantum wells~\cite{gunter, todorov_PRL2010, askenazi_NJoP2014_ultra-strong}.

The strong coupling regime can also be observed in the absence of optical confinement: when the coupling between a material oscillator and free space radiation is so strong that the spontaneous emission rate is 
comparable to the non-radiative damping rate, the optical resonance becomes radiatively broadened (as sketched in fig.~\ref{fig_introduction}b) and the properties of the system cannot be described using a perturbative treatment of 
the light-matter interaction~\cite{creatore}. In this work we show that ultra-strong coupling with free space radiation can be achieved when the matter excitation is a collective mode with extremely large oscillator strength and 
the radiative broadening is of the same order of magnitude as the matter excitation energy. Such a situation arises in very dense two-dimensional electron 
gases confined in semiconductor quantum wells (QWs). Indeed in these systems, dipole-dipole interaction among electronic transitions between confined levels of the quantum well (intersubband transitions) is responsible for the 
emergence of a bright collective mode of the electron gas, gathering the entire oscillator strength of the system~\cite{delteilPRL2012plasmons}. 
This collective mode, known as multisubband plasmon, has a superradiant nature~\cite{laurentPRL2015_superradiance}: radiative lifetimes as short as 10 fs have been reported, thus much shorter than any non-radiative 
scattering process in the structure. As the radiative broadening is larger than the non-radiative one, the collective excitation can be considered as strongly coupled with free space radiation: light -- matter interaction is a 
non-perturbative phenomenon, giving rise to mixed states. When the broadening becomes comparable to the oscillation frequency of 
the emitted mid-infrared radiation, the system enters the ultra-strong coupling regime and we show that, in that case, the RWA leads to unphysical predictions for the linear optical properties of the doped semiconductor layer, 
similarly to the microcavity case. Furthermore we show that Markov approximation, which is widely used in quantum optics, is not appropriate to describe ultra-strong coupling with free space radiation.
Our theoretical model uses the input - output formalism~\cite{ciuti_PRA2006_inpout_output} to solve the non-markovian equations for the dynamics of the coupled system. For an optical input, this formalism allows calculating 
the optical properties of the electron gas (transmissivity, reflectivity, absorptivity). For an input corresponding to the thermal fluctations of an electronic reservoir, we calculate the incandescent emission of the system and 
demonstrate that it follows Kirchhoff's law of thermal emission for a grey body~\cite{huppert_ACSph2015_Radiatively_broadened_incandescence}. In complement to the input-output description, we also perform the 
full diagonalization of the hamiltonian of the coupled system, thus describing the mixed light -- matter states and the associated Hopfield coefficients. 
This second approach provides the dispersion not only of the radiative, but also of the non-radiative (or localized) eigenmodes of the system.

The manuscript is organized as follows. In section \ref{sec_MSP}, we present our microscopic model of the collective excitations of the electron gas, the multisubband plasmons (MSPs). It is based on the dipole representation of the 
light-matter interaction in the Coulomb gauge\cite{todorovPRB2012plasmon_dipole_gauge, pegolottiPRB2014Multisubband_plasmons}. Section \ref{sec_H} provides the hamiltonian description of the full system which 
includes the MSP modes and their coupling to a photonic and an electronic reservoir. The time-evolution of this system is investigated in sections \ref{sec_QL} and \ref{sec_damp}, 
where we derive the input-output relations\cite{ciuti_PRA2006_inpout_output, walls_milburn2007quantum_optics, gardinerPRA1985_input_output}. In section \ref{sec_opt}, we apply this resolution method to the case of an 
optical input and compute the MSP absorptivity, transmissivity and reflectivity. 
The different regimes of light -- matter interaction are then analyzed, with particular attention to the ultra-strong coupling regime. 
We study the role of anti-resonant coupling terms and show that they influence significantly the spectral shape of the plasmon resonance when the radiative broadening is comparable with the plasmon energy, thus RWA is 
not appropriate to describe the ultra-strong coupling regime. We also show that Markov approximation is not accurate in this regime.
Section \ref{sec_incandescence} describes the MSP emission under an electronic excitation using the same input-output formalism.  Finally section~\ref{sec_GS} presents the diagonalization of the full light -- matter coupled Hamiltonian, together with an analysis of the properties of the new eigenstates of the system. Conclusions and perspectives of this work are drawn in section~\ref{sec:conclusion}. 

\section{Superradiant states in dense electron gases: multisubband plasmons}
\label{sec_MSP}
\begin{figure}
\centering  
\includegraphics[scale=0.8]{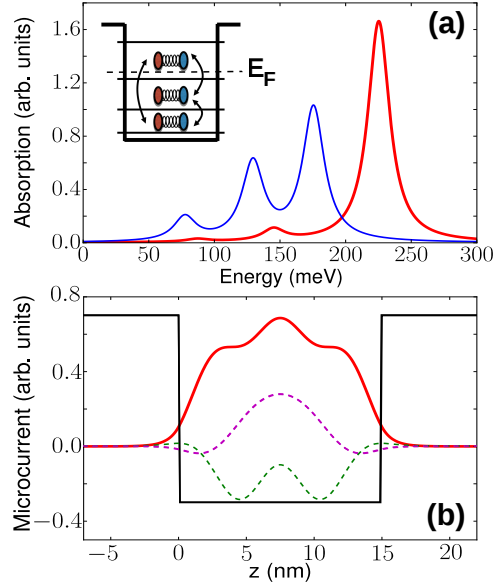}
\caption{(a) Low angle absorption of a 15 nm thick GaInAs/AlInAs quantum
well with electronic density $N_s=1.5\times10^{13}$ cm$^{-2}$. The
inset is a sketch of the interacting dipoles associated 
with the different intersubband transitions. The dashed line indicates the Fermi energy. (b) Plot of the current
densities associated with the three bright multisubband plasmons of the same
quantum well.}
\label{fig_plasmon}
\end{figure}

Our system is based on a highly doped semiconductor quantum well (QW) with several occupied subbands. 
Radiation polarized along the growth direction $z$ induces intersubband transitions, as sketched in the inset of fig.~\ref{fig_plasmon}a, where each transition is 
represented as an oscillator along $z$. In this system dipole -- dipole interaction between intersubband excitations is responsible for a strong modification of the optical spectrum~\cite{todorovPRB2012plasmon_dipole_gauge}, 
with the emergence of new bright collective modes, the multisubband plasmons~\cite{delteilPRL2012plasmons} (MSP).

In ref.~\onlinecite{pegolottiPRB2014Multisubband_plasmons} we have presented a quantum model allowing the calculation of the multisubband plasmon states in a highly 
doped quantum well. In this model the light -- matter coupling is treated by using the dipole representation of the Coulomb gauge~\cite{todorovPRB2012plasmon_dipole_gauge}, in which the interaction Hamiltonian is expressed as a 
function of the intersubband polarization density $\mathbf P$ and of the displacement field $\mathbf D$ as: 
\begin{equation}
\label{eq_H_plasmon}
H_{int}= \frac{1}{\epsilon_0 \epsilon_s}  \int \left[ \frac{1}{2}\mathbf P^2 (\mathbf r) - \mathbf D (\mathbf r) \cdot \mathbf P (\mathbf r) \right] d^3 \mathbf r = H_{dd} + H_I^{ph}
\end{equation}
with $\epsilon_s$ the material permittivity. The first term, $H_{dd}$, accounts for dipole -- dipole coupling between intersubband transitions, whereas the second one, $H_I^{ph}$, describes the coupling of the electronic excitations with the external field. As intersubband transitions are associated with a dipole oscillation along the growth direction $z$, only the components $P_z$ and $D_z$ contribute to the interaction Hamiltonian.

The intersubband polarization density operator is calculated as~\cite{todorovPRB2012plasmon_dipole_gauge}:
\begin{equation}
\hat{ P_z} (z, \mathbf r_{||})= \sum_{\alpha, \mathbf k} \frac{j_\alpha (z)}{w_\alpha} e^{i \mathbf k \cdot \mathbf r_{||} } 
\left( B^\dag_{\alpha, \mathbf k}+ B_{\alpha, -\mathbf k} \right)
\end{equation}
with $\alpha$ the intersubband transition index, 
$w_\alpha$ the intersubband transition frequency, and $\mathbf k$ the in-plane wavevector of the transition. The operators $B^\dag_{\alpha, \mathbf k}$, $B_{\alpha, \mathbf k}$ are bosonic creation and annihilation operators of the intersubband excitations. As only the transitions with low 
$\vert \mathbf k \vert$ are coupled to light, the electronic dispersion can safely be neglected and $w_{\alpha}$ is assumed independent of $\mathbf k$. 
The quantity $j_\alpha (z)$ is the intersubband current density and it is computed from the electronic wavefunctions together with the occupation of the corresponding subbands~\cite{pegolottiPRB2014Multisubband_plasmons}.

The multisubband plasmon Hamiltonian is given by~\cite{todorovPRB2012plasmon_dipole_gauge}: 
\begin{equation}
H_{pl}= \sum_{\alpha, \mathbf k} \hslash w_\alpha B^\dag_{\alpha, \mathbf k} B_{\alpha, \mathbf k} +H_{dd}
\end{equation}

As the dipole-dipole coupling induced by $H_{dd}$ is quadratic in the operators $B_{\alpha}$, the Hamiltonian $H_{pl}$ 
can be diagonalized through a Bogoliubov transformation to obtain its eigenmodes, the multisubband plasmons, characterized by the eigenfrequencies $\omega_n$ and the bosonic creation 
and annihilation operators $P_{n, \mathbf k}^\dag$ and $P_{n, \mathbf k}$, which are linear combinations of $B_{\alpha, \mathbf k}^\dag$ and $B_{\alpha, \mathbf k}$. 
The polarization then takes the form: 
\begin{equation}
\label{eq_polar_plasm}
\hat{ P_z} (z, \mathbf r_{||})= \sum_{n, \mathbf k} \frac{J_n (z)}{\omega_n} e^{i \mathbf k \cdot \mathbf r_{||} } \left( P^\dag_{n, \mathbf k}+ P_{n, -\mathbf k} \right)
\end{equation}
Here $J_n (z)$ are the current densities associated with the MSP eigenmodes, computed from the Bogoliubov coefficients. These quantities describe the spatial distribution of the collective electronic excitations and thus characterize the polarization of the medium, 
induced by the coupled intersubband transitions. 

As an example of application of our model, let us consider a 15 nm GaInAs/AlInAs quantum
well with electronic surface density $N_s=1.5\times10^{13}$ cm$^{-2}$. The absorption spectrum calculated in a single particle picture is represented by the blue line in fig.~\ref{fig_plasmon}a, displaying resonances at 
the energies of the optically active intersubband transitions. The red line presents the  absorption spectrum calculated by taking into account the dipole -- dipole interaction, showing that the coupling between intersubband 
transitions results in multisubband plasmon resonances at different energies with respect to the bare transitions. Figure~\ref{fig_plasmon}b presents the calculated MSP current densities. 
It can be shown~\cite{pegolottiPRB2014Multisubband_plasmons} that the effective oscillator strength of the MSPs, i.e. the absorption amplitude, is proportional to 
$\left \vert \int{dz J_n(z)} \right \vert ^2$.  Panels (a) and (b) show that, in agreement with experimental observations, a main bright mode concentrates almost the entire oscillator strength of the system.

\begin{figure}
\centering  
\includegraphics[scale=0.8]{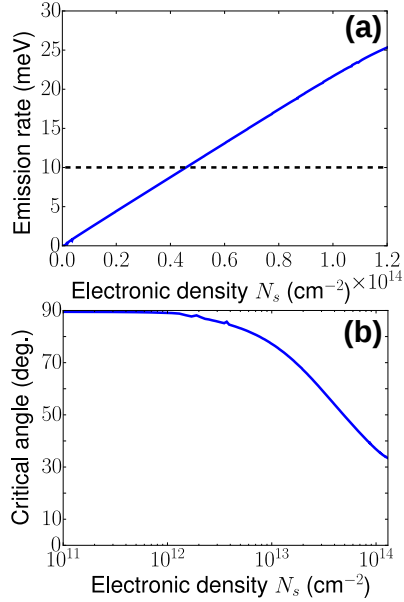}
\caption{
(a) Spontaneous emission rate $\hslash \Gamma_0 $ of the main bright multisubband plasmon of a 100 nm wide GaInAs/AlInAs quantum well, plotted as a function of the electronic density in the quantum well $N_s$. 
The dashed line shows the typical value of the plasmon non-radiative decay rate $\hslash \gamma$. (b) Plot of the critical angle such that 
$\hslash \Gamma(\theta)=\hslash \gamma$ as a function of $N_s$ for the same quantum well.}
\label{fig_CF}
\end{figure}

The spontaneous emission rate of the MSP is also proportional to $\left \vert \int{dz J_n(z)} \right \vert ^2$. For the main bright multisubband plasmon, at frequency $\omega_0$, the spontaneous emission rate calculated by using Fermi's golden rule, is given by: 
\begin{equation}
\Gamma (\omega_0, \theta)  =\Gamma_0 \frac{\sin^2 \theta}{\cos \theta} \ \text{with } \Gamma_0= \frac{S}{\hslash \epsilon_0 \sqrt{\epsilon_s}} \frac{|\int dz \ J_0(z)|^2}{c \, \omega_0} \propto N_s \label{eq_FGR} 
\end{equation}
In this formula, $S$ is the system area, $\theta$ is the emission angle of the bright plasmon, determined by its frequency $\omega_0$ and its in-plane wavevector $\mathbf k$ according to $ \sin \theta = \frac{ck}{\sqrt{\epsilon_s}\omega_0}$. 
The expression for $\Gamma_0$ is derived in appendix \ref{sec_app_lamb}. Note that the current density $J_0$ is proportional to $S^{-\frac{1}{2}}$ so that $\Gamma_0$ is independent of the system 
area\cite{pegolottiPRB2014Multisubband_plasmons} (as long as the dimensions of the system are large compared to the optical wavelength). Figure~\ref{fig_CF}a presents the spontaneous emission rate $\hslash \Gamma_0$ calculated for a 100~nm 
GaInAs/AlInAs quantum well, as a function of the electronic density per unit surface $N_s$. The spontaneous emission rate is approximately proportional to $N_s$ and it can reach several tens of meV, which is larger than the typical non-radiative broadening of the plasmon resonance, 
$\hslash \gamma \simeq 10$ meV. This was experimentally demonstrated in ref.~\onlinecite{laurentPRL2015_superradiance} in a highly doped GaInAs quantum well, where a spontaneous emission time as short as few tens of femtosecond 
was measured. This is much shorter than any non-radiative scattering event and therefore plasmons are in the strong coupling regime, where their relaxation dynamics is dominated 
by the radiative rate. In the following sections we show that in this regime the interaction of the MSP with free space radiation cannot be treated perturbatively and that the absorptivity 
of the quantum well ceases to be proportional to $N_s$.

In addition, the emission rate in equation \eqref{eq_FGR} depends strongly on the angle $\theta$ and diverges at 90$^\circ$. Therefore any QW system reaches the strong coupling regime for large enough $\theta$. 
This is illustrated in figure~\ref{fig_plasmon}b, which shows the density dependence of the angle $\theta$ for which the strong coupling condition $\hslash \Gamma(\omega_0, \theta)=\hslash \gamma$ is fulfilled in a GaInAs/AlInAs quantum well. It can be seen 
that up to $N_s=10^{12}$ cm$^{-2}$ (i.e. in the case of usual intersubband devices), the critical angle is extremely close to 90$^\circ$, and the plasmons can be considered in the weak coupling regime. 
However for higher doping levels, the critical angle decreases, reaching 40$^\circ$ at $N_s=10^{14}$ cm$^{-2}$. In that case the MSPs are in strong coupling with the free space radiation for most angles $\theta$, 
and light-plasmon interaction must be described non-perturbatively. 

At high densities $N_s$ and high angles $\theta$, the bright plasmon even reaches the regime of ultra-strong coupling with free space radiation when $\Gamma (\omega_0, \theta)$ is comparable with the bare frequency $\omega_0$. 
In that regime, the frequency dependence of the emission rate must be taken into account, and, even in the absence of a photonic confinement, the anti-resonant terms of the light-matter interaction 
modify substantially the lineshape of the plasmon resonance and are necessary to ensure that MSP reflectivity is suppressed at high frequencies, as it will be discussed in the following. 

\section{Input -- output theory of superradiant states}
\label{sec_model}
The dependence of the MSP spontaneous emission rate on the light propagation direction and on the electronic density allows observing three different light-matter coupling 
regimes: weak, strong and ultra-strong coupling. In order to describe within the same theoretical framework the three regimes, it is necessary to treat the interaction between the electromagnetic field 
and the MSP in a non-perturbative way, including the anti-resonant terms. Furthermore, it is important to take into account both radiative and non-radiative broadening of the plasmon.

These two requirements are fulfilled by our quantum model, presented in this section. It is based on the resolution of time evolution equations in the input -- output formalism. When considering an optical input 
(section \ref{sec_opt}), we describe the optical properties of the MSP (reflectivity, absorptivity, transmissivity). In the presence of an electronic input (section \ref{sec_incandescence}), our model allows calculating the 
MSP incandescent emission. Note that, even though the model is applied to the study of MSP properties, its results, expressed in terms of normalized quantities, are general and also apply to other systems with radiatively broadened 
excitations.   
\subsection{Coupled system Hamiltonian}
\label{sec_H}

\begin{figure}
\centering  
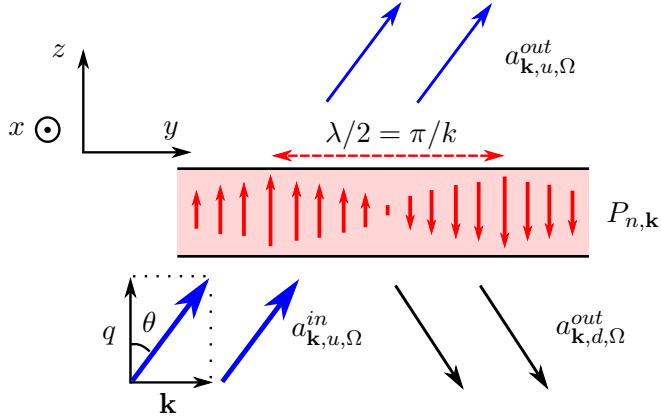  
\caption{Schematic representation of the multisubband
excitation of a doped quantum well, together with the incoming and
outgoing optical radiation and associated operators.}
\label{schema}
\end{figure}

The multisubband plasmons are coupled with two reservoirs: the free electromagnetic field in the dielectric medium and a bath of electronic excitations responsible for the non-radiative broadening of MSPs. 
The complete Hamiltonian of the system is composed of: 
$$
H=H_{pl}+H_{el}+H_{ph} + H_I^{el}+ H_I^{ph},
$$
with the isolated system Hamiltonians: 
\begin{eqnarray*}
H_{pl}+H_{el}+H_{ph} &=& \sum_{n, \mathbf k} \hslash \omega_{n, \mathbf k} P^\dag_{n, \mathbf k} P_{n, \mathbf k} 
+ \sum_{n, \mathbf k} \int d\Omega \  \hslash \Omega \ b^\dag_{n, \mathbf k, \Omega } b_{n, \mathbf k, \Omega}\\
&+& \sum_{\mathbf k}\int d\Omega \  \hslash \Omega \  a_{\mathbf k, u, \Omega}^\dag a_{\mathbf k, u, \Omega} +\sum_{\mathbf k}\int d\Omega \  \hslash \Omega \  a_{\mathbf k, d, \Omega}^\dag a_{\mathbf k, d, \Omega} 
\end{eqnarray*}
The operators $a_{ \mathbf k, u, \Omega}$ and $a_{ \mathbf k, d, \Omega}$ describe respectively the continuum of upward and downward propagating photon modes of the electromagnetic environment of the well. 
We denote $\Omega$ the photon frequency, such that $\Omega=\frac{c}{\sqrt{\epsilon_s}} \sqrt{k^2+q^2}$, where $\mathbf k$ and $q$ are respectively the in-plane and growth axis components of the photon 
wavevector (see fig.~\ref{schema}) and $\epsilon_s$ is the dielectric constant 
of the embedding semiconductor. Note that only transverse magnetic (TM) polarized modes are considered since transverse electric modes do not couple with the MSP polarization, which is oriented along the growth axis $z$. 
The charge density wave associated to the MSP, with a dipole oriented along the growth direction, is sketched in figure~\ref{schema} (red arrows).

The light-matter interaction $H_I^{ph}$ characterizes the coupling between the MSP modes and the photonic reservoir. As the system is invariant under translation along the quantum well plane, this interaction conserves $\mathbf k$, 
and each MSP is coupled to all photon modes with the same in-plane wavevector so that $H_I^{ph}$ takes the form: 
\begin{eqnarray}
H_I^{ph} &=& i \hslash \sum_{\mathbf k, n, s} \int d\Omega \  W_{n, \mathbf k,  \Omega} \left[ a_{\mathbf k, s, \Omega}^\dag -  a_{-\mathbf k, s, \Omega} \right] 
\left[ P_{n,\mathbf k}+ P_{n,-\mathbf k}^\dag   \right]  \label{eq_interaction}
\end{eqnarray}
where $s$ labels the upwards and downwards photon modes and $W_{n, \mathbf k,  \Omega}$ is the coupling constant between the plasmon $n$ and the photon with frequency $\Omega$ and wavevector $\mathbf k$. 
$W_{n, \mathbf k,  \Omega}$, whose expression is provided in appendix \ref{sec_app_lamb}, is proportional to the integrated MSP current density: $W_{n, \mathbf k,  \Omega} \propto |\int dz \ J_n(z)| $. 
Hamiltonian \eqref{eq_interaction} includes not only resonant terms, i.e. products of an annihilation and a creation operator, which describe photon emission and absorption, but also anti-resonant terms, involving 
two annihilation or two creation operators. The last will be shown to play a determinant role in the optical response of highly doped QWs. 
\begin{figure}
\centering  
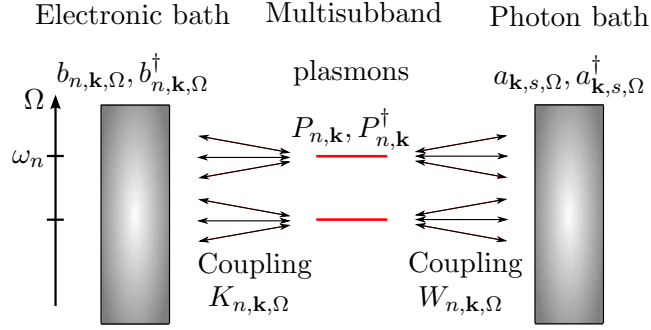  
\caption{Schematic representation of the MSP discrete modes, and of the electronic and photonic bath, labeled by the corresponding operators and coupling constants.}
\label{fig_system}
\end{figure}
 
The bosonic operators $b_{n, \mathbf k, \Omega}$ describe a reservoir of electronic excitations responsible for non-radiative relaxation and excitation of MSPs. Similarly to light-matter interaction, the coupling between plasmons and the electronic bath is written:
\begin{eqnarray*}
H_I^{el}&=& i \hslash \sum_{n,\mathbf k} \int d\Omega \  \left[ K_{n,\mathbf k, \Omega} b_{n,\mathbf k, \Omega}^\dag -K_{n,\mathbf k, \Omega}^* b_{n,-\mathbf k, \Omega} \right]
 \left[ P_{n,\mathbf k}+ P_{n,-\mathbf k}^\dag   \right] 
\end{eqnarray*}
The coupling constants $K_{n,\mathbf k, \Omega} $ will be specified later. Figure~\ref{fig_system} provides a schematic representation of the multisubband plasmon states and of their coupling with the two bosonic reservoirs.

In the following, we focus on the optical properties of highly doped structures with a single bright MSP, as discussed in relation with fig.~\ref{fig_plasmon} and experimentally observed in ref.~\onlinecite{delteilPRL2012plasmons}. Its eigenfrequency is denoted $\omega_0$ and the coupling of other plasmons with the electromagnetic 
field is neglected.  

\subsection{Time evolution equations}
\label{sec_QL}
In the Heisenberg representation, the time-evolution of the bath and plasmon operators under the Hamiltonian $H$ is given by the following equations: 
\begin{eqnarray}
\dot a_{ \mathbf k, s, \Omega} &=& -i \Omega \  a_{ \mathbf k, s, \Omega} + W_{ \mathbf k, \Omega} \left[P_{ \mathbf k}  + P^\dag_{ -\mathbf k}  \right] \label{var_a_anti} \\
\dot b_{ \mathbf k, \Omega} &=& -i \Omega \  b_{ \mathbf k, \Omega} + K_{ \mathbf k, \Omega} \left[P_{ \mathbf k}  + P^\dag_{ -\mathbf k}  \right] \label{var_b_anti} \\
\dot P_{ \mathbf k} &=& -i \omega_{0} P_{ \mathbf k} + \int d\Omega \  \left[K_{-\mathbf k, \Omega}  b^\dag_{- \mathbf k, \Omega} - K_{ \mathbf k, \Omega}^* b_{ \mathbf k, \Omega}  \right] \nonumber \\
&& + \sum_s \int d\Omega \  \left[W_{-\mathbf k, \Omega}  a^\dag_{- \mathbf k, s, \Omega}-W_{ \mathbf k, \Omega} a_{\mathbf k, s, \Omega} \right] \label{var_P_anti} 
\end{eqnarray}
We introduce initial time $t_0$ and final time $t$ and define the input and output operators:
\begin{eqnarray*}
a_{ \mathbf k, s, \Omega}^{in}=a_{\mathbf k, s, \Omega}(t_0)\  e^{i \Omega t_0} \ &\text{and}& \ b_{\mathbf k, \Omega}^{in}=b_{ \mathbf k, \Omega}(t_0) \  e^{i \Omega t_0} \\
a_{ \mathbf k, s, \Omega}^{out}=a_{\mathbf k, s, \Omega}(t) \  e^{i \Omega t} \ &\text{and}& \ b_{\mathbf k, \Omega}^{out}=b_{ \mathbf k, \Omega}(t) \  e^{i \Omega t} \\
\end{eqnarray*} 
In the limit $t_0 \rightarrow - \infty$, and $t \rightarrow + \infty$, equations \eqref{var_a_anti} and \eqref{var_b_anti} yield the relation between the output fields $a^{out}$ and $b^{out}$ and the input operators $a^{in}$ and $b^{in}$: 
\begin{eqnarray}
&&b_{ \mathbf k, \Omega}^{out} = b_{ \mathbf k, \Omega}^{in}+ K_{ \mathbf k, \Omega} 
\left[ \tilde P_{ \mathbf k} (\Omega) + \tilde P_{ -\mathbf k}^\dag (\Omega) \right] \label{var_b_out} \\
&&a_{\mathbf k, s, \Omega}^{out} = a_{ \mathbf k, s, \Omega}^{in}+ W_{ \mathbf k, \Omega} 
\left[ \tilde P_{ \mathbf k} (\Omega) + \tilde P_{ -\mathbf k}^\dag (\Omega) \right] \label{var_a_out} 
\end{eqnarray}
where $\tilde P_{ \mathbf k} (\omega)$ and $\tilde P_{ -\mathbf k}^\dag (\omega)$ are the Fourier transform of the operators $P_{ \mathbf k}(t)$ and $P_{ -\mathbf k}^\dag (t)$, respectively. 
The creation and annihilation plasmon operators are related by the following formula, obtained by taking the hermitian conjugate of equation \eqref{var_P_anti}:
\begin{equation}
 \tilde P^\dag_{- \mathbf k} (\omega) = \frac{\omega_{0}-\omega}{\omega_{0}+\omega}  \tilde P_{\mathbf k} (\omega) \label{crea_anni}
\end{equation}
By Fourier transforming equation \eqref{var_P_anti} we also get: 
\begin{equation}
 \tilde P_{\mathbf k}(\omega)=\frac{\tilde F_{\mathbf k}(\omega)}
{i(\omega-\omega_0) -  \frac{2\omega_{0}}{\omega_{0}+\omega} \left[\frac{\tilde \Gamma^{}_{\mathbf k} (\omega) }{2}
 - \frac{\tilde \Gamma^{*}_{\mathbf k} (-\omega)}{2} \right] }. \label{eq_M}
\end{equation}
The factor $\frac{2\omega_{0}}{\omega_{0}+\omega}$ in this expression is due to anti-resonant coupling terms. It is unity for 
$\omega = \omega_0$ but it becomes significant far from the resonance, thus profoundly affecting the optical response of highly doped QWs, as it will be discussed in section \ref{sec_opt}.  

The damping kernel $\Gamma_{\mathbf k}=\Gamma_{\mathbf k}^{ph}+\Gamma_{\mathbf k}^{el}$ is defined in time domain as a function of the delay $\tau$:
\begin{eqnarray}
\Gamma^{el}_{\mathbf k} (\tau)&=&  2 \Theta (\tau) \int d\Omega \ |K_{\mathbf k, \Omega}|^2 \  e^{-i\Omega \tau}    \label{gam_el_t}\\
\Gamma^{ph}_{\mathbf k} (\tau)&=& 2 \Theta (\tau)  \int d\Omega \  2  W_{\mathbf k, \Omega}^2 \  e^{-i\Omega \tau}  \label{gam_ph_t}
\end{eqnarray}
where $\Theta (\tau)$ is the Heaviside function. The additional factor 2 in expression \eqref{gam_ph_t} is due to the presence of upward and downward propagating modes which both contribute to the plasmon radiative damping. 
 
The function $\tilde F_{\mathbf k}= \tilde F^{el}_{\mathbf k} + \tilde F^{ph}_{\mathbf k, u} +\tilde F^{ph}_{\mathbf k, d} $ is the driving force defined as:
\begin{eqnarray}
\tilde F^{el}_{\mathbf k} (\Omega)&=&   2\pi \  K_{\mathbf k, \Omega}^*  \  b_{\mathbf k, \Omega}^{in} \label{eq_F_el} \\
\tilde F^{ph}_{\mathbf k, s} (\Omega)&=& 2\pi \  W_{\mathbf k, \Omega} \  a_{\mathbf k, s, \Omega}^{in} \label{eq_F_ph}
\end{eqnarray}
Note that $\tilde F_{\mathbf k}$ is zero for negative $\Omega$, as all the bath operators have positive energy. Indeed, due to the presence of antiresonant terms, including negative-frequency contributions would yield unphysical 
results, with non-zero output even for a vacuum input.

Combining equations \eqref{var_b_out} and \eqref{var_a_out} with \eqref{eq_M}, \eqref{eq_F_el} and \eqref{eq_F_ph}, we derive a linear relation between the input and output fields: 
\begin{equation}
 \label{eq_U}
 \left( \begin{array}{c}
a_{ \mathbf k, u, \Omega}^{out} \\
a_{ \mathbf k, d, \Omega}^{out} \\
b_{\mathbf k,\Omega}^{out}
\end{array} \right)
= U (\mathbf k, \Omega) 
 \left( \begin{array}{c}
a_{ \mathbf k, u, \Omega}^{in} \\
a_{ \mathbf k, d, \Omega}^{in} \\
b_{ \mathbf k,\Omega}^{in} 
\end{array} \right)
\end{equation}
The input-output matrix $U(\mathbf k, \Omega)$ is a unitary matrix whose coefficients are specified in the following. It is a generalized scattering matrix of the multisubband plasmon, describing the output corresponding to an 
optical or to an electronic input. The unitarity of the matrix ensures the following energy conservation relation:
\begin{equation}
 \langle  a_{ \mathbf k, u, \Omega}^{out\dag} a_{ \mathbf k, u, \Omega}^{out} + a_{ \mathbf k, d, \Omega}^{out\dag} a_{ \mathbf k, d, \Omega}^{out}  + b_{ \mathbf k,\Omega}^{out\dag} b_{ \mathbf k,\Omega}^{out}\rangle =
  \langle  a_{ \mathbf k, u, \Omega}^{in \dag} a_{ \mathbf k, u, \Omega}^{in} + a_{ \mathbf k, d, \Omega}^{in \dag} a_{ \mathbf k, d, \Omega}^{in}  + b_{ \mathbf k,\Omega}^{in \dag} b_{ \mathbf k,\Omega}^{in}\rangle
\end{equation}

\subsection{Radiative and non-radiative damping}
\label{sec_damp}
Radiative and non-radiative damping functions $ \tilde \Gamma^{ph}_{\mathbf k } $ and $\tilde \Gamma^{el}_{\mathbf k } $ are evaluated by Fourier transforming relations \eqref{gam_el_t} and \eqref{gam_ph_t}. 
The real part of the photon damping function is given by (see appendix \ref{sec_app_lamb}): 
\begin{eqnarray}
\text{Re} \left[ \tilde \Gamma^{ph}_{\mathbf k } (\omega) \right]  &=& 2 \int d\Omega \ 2  W_{\mathbf k, \Omega}^2 \  \pi \ \delta(\omega-\Omega ) \nonumber \\
&=& 4\pi \ W_{ \mathbf k, \omega}^2   \label{realGam} \\
&=&  \Gamma_0 \frac{ ck^2}{\sqrt{\epsilon_s} \omega_0 \sqrt{\frac{\epsilon_s \omega^2}{c^2}- k^2}} \Theta \left(\frac{\epsilon_s \omega^2}{c^2}- k^2 \right) \label{realGammafinal}
\end{eqnarray}

It characterizes the decay of the MSP mode due to its interaction with the photonic bath. 
Optical experiments are usually performed at a fixed light propagation angle $\theta$, such that $ \sin \theta = \frac{ck}{\sqrt{\epsilon_s}\Omega}$ (see fig. \ref{schema}). It is thus useful to write the real part of 
the photon damping function as:
\begin{equation}
\Gamma (\theta,\omega) =\Gamma_0 \frac{\omega}{\omega_0} \frac{\sin^2 \theta}{\cos \theta}=\Gamma(\theta, \omega_0) \frac{\omega}{\omega_0}
\label{eq_Gam_th} 
\end{equation}
The emission rate $\Gamma (\theta,\omega)$ is proportional to that obtained using Fermi's golden rule and provided in equation \eqref{eq_FGR}. However in the input-output approach, the radiative decay rate is also 
frequency-dependent: this dependence is characteristic of non-markovian dynamics and it will be shown to be determinant in the ultra-strong coupling regime. Only at resonance, $\omega=\omega_0$, does the emission rate $\Gamma(\theta, \omega)$ coincide with the result obtained by Fermi's golden rule.
Equation \eqref{realGammafinal} also shows that the real part of $\tilde \Gamma^{ph*}_{\mathbf k }(-\omega)$ is zero. 

The imaginary part $\text{Im}\left[  \tilde \Gamma^{ph}_{\mathbf k }\right]$ corresponds to a shift of the plasmon frequency, known in atomic physics as Lamb shift~\cite{cohen_atom_photon}. 
However, it is shown in appendix \ref{sec_app_lamb} that the imaginary parts of $\tilde \Gamma^{ph}_{\mathbf k }(\omega) $ and 
$\tilde \Gamma^{ph*}_{\mathbf k }(-\omega)$ are equal, so that the Lamb shift is zero for the system under study. Therefore, only the real part of the damping function 
$\text{Re}\left[  \tilde \Gamma^{ph}_{\mathbf k }(\omega)\right] = \Gamma (\theta,\omega)$ has to be considered in equation \eqref{eq_M}. The vanishing Lamb shift is a peculiar consequence of the dependence on 
$\Omega$ of the coupling constant $W_{\mathbf k, \Omega}$. This dependence is characteristic of the coupling between a two-dimensional excitation polarized along $z$, and the three-dimensional electromagnetic radiation. 
Note that the vanishing Lamb shift is also obtained as a consequence of the long-wavelength approximation, $q\ll L^{-1}$ (with $L$ the QW width). For thick heterostructures, e.g. containing several QWs, the shift might be non-zero 
and it might even be enhanced by the superradiance effect~\cite{rohlsberger, scully}.

As for the electronic damping, the imaginary part of $\tilde \Gamma_{\mathbf k}^{el}$ is neglected, and its real part is approximated by a constant, according to Markov first approximation~\cite{gardinerPRA1985_input_output}. More precisely, 
we use the following expression:
\begin{equation}
\tilde \Gamma^{el}_{\mathbf k } (\omega)   = 2 \pi \left|K_{ \mathbf k, \omega} \right|^2   \equiv \gamma \  \Theta (\omega) \label{defgam} 
\end{equation} 
In this equation the Heaviside function $\Theta$ must be included because considering excitations of the electronic bath with negative frequency $\omega$ would be unphysical\cite{ciuti_PRA2006_inpout_output}. Indeed, due to 
the presence of antiresonant coupling terms, setting simply $\Gamma^{el}_{\mathbf k } (\omega)=\gamma$, as in usual Markov approximation~\cite{gardinerPRA1985_input_output}, suppresses the effect of non-radiative damping because of 
spurious compensation between resonant and anti-resonant plasmon decay processes. 

\section{Optical properties of superradiant modes}
\label{sec_opt}

The optical scattering matrix provides the output (or scattered) radiation corresponding to a given input (or incident) radiation. It is given by the upper-left block of the input-output matrix $U$: 
\begin{eqnarray*}
t_{\mathbf k}(\Omega)&=&  U_1^1(\mathbf k ,\Omega)= U_2^2(\mathbf k ,\Omega)\\ 
r_{\mathbf k}(\Omega)&=&  U_2^1(\mathbf k ,\Omega) =U_1^2(\mathbf k ,\Omega) 
\end{eqnarray*}
According to the analysis of the previous section, the angle-dependent transmission and reflection amplitudes are given by:
\begin{eqnarray}
t (\theta, \omega) &=& 
\frac{i(\omega-\omega_0) - \frac{2\omega_0}{\omega_0 + \omega} \frac{\gamma}{2}}{i(\omega -\omega_0) - \frac{2\omega_0}{\omega_0 + \omega}\left[\frac{\gamma}{2}+\frac{\Gamma (\theta,\omega)}{2} \right]} \label{eq_t}\\
r (\theta, \omega) &=& 
\frac{\frac{2\omega_0}{\omega_0 + \omega} \frac{\Gamma (\theta,\omega)}{2}}{i(\omega -\omega_0) - \frac{2\omega_0}{\omega_0 + \omega}\left[\frac{\gamma}{2}+\frac{\Gamma (\theta,\omega)}{2} \right]} \label{eq_r} \label{eq_rt} 
\end{eqnarray}
From these formulae, we can derive the absorptivity of the QW:
\begin{eqnarray}
\alpha (\theta, \omega) &=& 1- |t (\theta, \omega)|^2-|r (\theta, \omega)|^2 \nonumber \\
&=& \frac{\frac{4\omega_0^2}{(\omega_0 + \omega)^2} \frac{\gamma \Gamma (\theta,\omega)}{2}}{(\omega-\omega_0)^2 + \frac{4\omega_0^2}{(\omega_0 + \omega)^2}\left[\frac{\gamma}{2}+\frac{\Gamma (\theta,\omega)}{2} \right]^2} 
\label{eq_alpha} 
\end{eqnarray}
It is important to underline that the expressions obtained for $r$ and $t$ are identical to the one derived in Ref.~\onlinecite{alpeggianiPRB2014semiclassical_plasmons}, using a semiclassical description of the MSP optical 
properties based on a non-local susceptibility formalism. 
\begin{figure}[h]
\centering  
\includegraphics[scale=0.9]{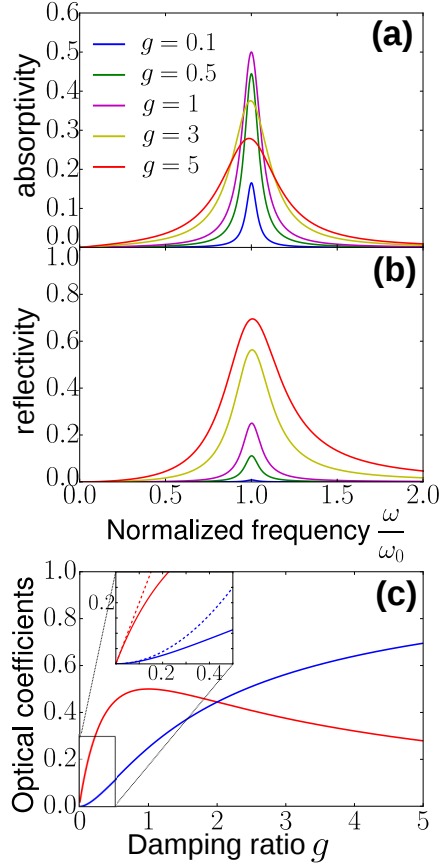}
\caption{(a) Plot of the absorptivity  $\alpha (\theta, \omega)$ as a function of the normalized frequency $\frac{\omega}{\omega_0}$ for different values of the ratio between radiative and non-radiative damping rates 
$g=\frac{\Gamma(\theta, \omega_0)}{\gamma}$. (b) Reflectivity $|r (\theta, \omega)|^2$ spectra calculated for different values of $g$. (c)
Absorptivity (red) and reflectivity (blue) peak values as a function of $g$. Dashed lines correspond to the perturbative results valid in the weak coupling regime. All
spectra are obtained for $Q=15$.}
\label{fig_coeffs}
\end{figure}

The angle-dependent reflection and transmission amplitudes and the absorptivity can be recast as functions of three dimensionless parameters: 
the dimensionless frequency $\frac{\omega}{\omega_0}$, the damping ratio $g = \frac{\Gamma (\theta,\omega_0)}{\gamma}$ and the quality factor $Q=\frac{\omega_0}{\gamma}$. 
The value of these two last parameters is 
crucial as it determines the light--matter coupling regime in which the system stands. In the following we distinguish three characteristic regimes associated with very different optical properties: 
weak, strong and ultra-strong coupling.
Note that the factor $g$ is approximately proportional to $N_s$, as shown in section \ref{sec_MSP}, and for a given electronic density it also depends strongly on the incidence angle $\theta$, increasing 
from zero at normal incidence to infinity for optical excitation parallel to the QW plane. Therefore the three different regimes can be experimentally studied by either using quantum wells with different doping densities, or on a single sample, by changing the incidence angle $\theta$ . 

Figure \ref{fig_coeffs}(a) and (b) present respectively the absorptivity $\alpha (\theta, \omega)$ and the reflectivity $|r(\theta, \omega)|^2$ plotted as a function of the normalized frequency $\omega/\omega_0$, 
calculated for different values of $g$.  Clearly the damping ratio $g$ controls the shape of the spectra, as the linewidth of the plasmon resonance increases steadily with $g$ due to radiative broadening, 
as well as their amplitude. This can be ascribed to the fact that absorption and reflection are twofold 
processes: the first step is the creation of a plasmon through photon absorption with radiative rate $\Gamma(\theta, \omega)$; the second step is the plasmon decay into the electronic reservoir with a rate $\gamma$ in the case of 
absorption, or back into the photonic reservoir (with rate $\Gamma(\theta, \omega)$) in the case of reflection. The ratio $g$ determines the dominant decay mechanism and therefore fixes the relative importance of 
reflection and absorption processes.

Figure~\ref{fig_coeffs}(c) summarizes the peak values of the absorptivity (red continuous line) and reflectivity (blue continuous line) as a function of the damping ratio $g$. In the weak coupling regime, i.e. when $g \ll 1$ (see inset), 
MSPs decay mostly through the non-radiative channel and reflection is largely dominated by absorption. The absorptivity and reflectivity peaks have a Lorentzian lineshape of width 
$\frac{\gamma}{\omega_0}$ and respective height $2 g$ and $g^2$, as expected from a perturbative treatment of the light matter-interaction with fixed non-radiative broadening $\gamma$. Absorptivity and reflectivity calculated in this 
approximation are plotted as dashed lines in the inset of figure~\ref{fig_coeffs}(c), showing a very good agreement with the complete quantum model up to $g=0.1$. This is the situation usually encountered for quantum wells employed in mid-infrared 
optoelectronic devices (like quantum cascade lasers or quantum well infrared photodetectors), where absorptivity is proportional to the electronic density. As an example, for a GaInAs quantum well with $N_s=5 \times 10^{11}$ cm$^{-2}$, and a typical non-radiative broadening $\hbar \gamma=10$~meV for a mid-infrared resonance, at Brewster angle $\theta=17^\circ$ one gets $g\simeq 10^{-3}$. 

For $g >0.1$ the perturbative approach becomes less accurate: the system enters the strong coupling regime. Absorptivity and reflectivity peak values increase with $g$ at a smaller rate than expected 
from perturbation theory (see fig.~\ref{fig_coeffs}(c)), while 
the resonance linewidth is progressively broadened due to the increase of the radiative decay rate.

While the peak value of the reflectivity monotonously increases with $g$ towards one when $g\rightarrow \infty$, the absorptivity reaches its maximum value of 0.5 at $g=1$, then decreases towards zero. 
Indeed at $g=1$, critical coupling is achieved: plasmon radiative and non-radiative decay take place with the same characteristic rate. The width of the absorptivity peak is then 
$2\frac{\gamma}{\omega_0}$. Note that, if the QW is placed close to a metallic mirror, the transmissivity becomes zero at $\omega=\omega_0$ while the absorptivity is unity. 
The system thus displays total absorption at the MSP 
frequency\cite{huppert_ACSph2015_Radiatively_broadened_incandescence}. As shown in fig. \ref{fig_CF}(b), in a GaInAs quantum well for densities up to $N_s=10^{12}$ cm$^{-2}$ the critical coupling condition is only reached for $\theta \simeq 90^\circ$, and the MSP 
is in weak coupling with the free space radiation for almost all angles.

For $g > 10$, the plasmonic decay is mostly radiative, reflectivity monotonously increases with $g$, while the peak absorptivity decreases as $\frac{2}{g}$. 
In this case, the absorptivity at resonance is thus inversely proportional to the electronic density $N_s$. This counter-intuitive behavior is in remarkable contrast with the weak coupling limit. Furthermore, reflection 
becomes dominant over absorption, as shown in fig.~\ref{fig_coeffs}, and the transmissivity at resonance falls approximately to zero, corresponding to a metallic behavior of the doped layer. 
\begin{figure}
\centering  
\includegraphics[scale=0.9]{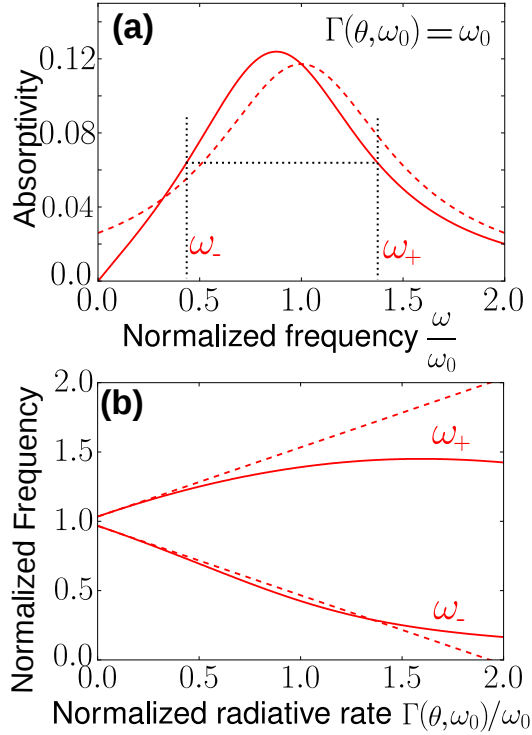}
\caption{(a) Plot of the plasmon absorptivity $\alpha (\theta, \omega)$ (red continuous line) as a function of the normalized frequency $\frac{\omega}{\omega_0}$. 
The dashed line results from applying Markov and rotating-wave approximations to the interaction with the photon reservoir. Both spectra are obtained for $Q=g=15$, i.e. $\Gamma(\theta, \omega_0)=\omega_0$.
(b) Plot of the normalized frequencies at half the maximum of the absorptivity $\frac{\omega_\pm}{\omega_0}$, as a function of the ratio $\frac{\Gamma(\theta,\omega_0)}{\omega_0}$. 
The dashed lines show the values $1\pm \left[\frac{\gamma}{2\omega_0}+\frac{\Gamma(\theta,\omega_0)}{2\omega_0}\right]$ expected from Markov approximation. 
The approximation becomes inaccurate when $\Gamma(\theta,\omega_0)$ is a significant fraction of $\omega_0$. This deviation is characteristic of the ultra-strong coupling regime. }
\label{fig_ultra1}
\end{figure}

We now investigate the case where not only $g \gg 1$, but $\Gamma(\theta, \omega_0)$ becomes comparable to $\omega_0$ ($Q \sim g$). We demonstrate that in this regime, corresponding to MSPs ultra-strongly coupled with free space 
radiation, the absorption spectra deviate significantly from a Lorentzian function. Furthermore anti-resonant terms of the interaction are crucial in preventing an unphysical high frequency behavior of the reflectivity. 

The dashed line in fig.~\ref{fig_ultra1}(a) presents the absorptivity spectrum obtained applying Markov and rotating-wave approximations to the light-matter coupling, for $\Gamma(\theta, \omega_0)=\omega_0$ i.e. in the ultra-strong 
coupling regime. It is a Lorentzian function centered at $\omega_0$ with full width at half the maximum (FWHM) $\omega_+ - \omega_- = \gamma + \Gamma(\theta,\omega_0)$. This spectrum is obviously unphysical as the absorptivity tends to a constant 
value for $\omega \rightarrow 0$. This unrealistic behavior is corrected within the full quantum model (continuous line) which does not rely on Markov approximation for the photonic reservoir: the linear dependence of 
$\Gamma(\theta,\omega)$ on the frequency ensures that both $r(\theta, \omega)$ and $\alpha(\theta,\omega)$ vanish for $\omega \rightarrow 0$. Figure~\ref{fig_ultra1}(b) displays a further representation of the same effect. 
The frequencies at half the maximum $\omega_+$ and $\omega_-$ (continuous lines) are plotted as a function of the normalized radiative rate $\Gamma(\theta, \omega_0)/\omega_0$. They are compared to the frequencies at half the maximum 
obtained within Markov approximation: $\frac{\omega_\pm}{\omega_0}=1 \pm \left[ \frac{\gamma}{\omega_0}+\frac{\Gamma(\theta, \omega_0)}{\omega_0} \right]$. 
Above $\frac{\Gamma(\theta, \omega_0)}{\omega_0}=g/Q=0.3$, that approximation is not accurate anymore, the system enters the ultra-strong coupling regime and $\omega_\pm$ deviate from their 
linear dependence. Note that the factor $\omega/\omega_0$ in $\Gamma(\theta, \omega)$ prevents reaching the unphysical value $\omega_-=0$ for any finite ratio $\frac{\Gamma(\theta, \omega_0)}{\omega_0}$. 
This is reminiscent of the no-go theorem for an electron gas coupled with a microcavity mode~\cite{nataf_naturecomm}.

\begin{figure}
\centering  
\includegraphics[scale=0.9]{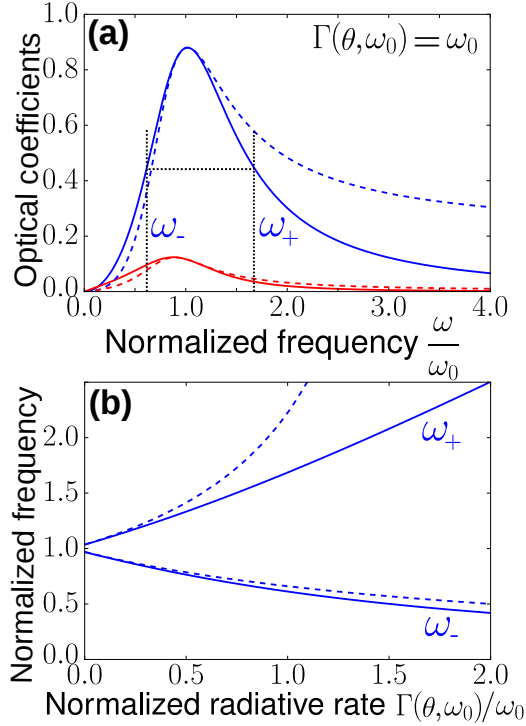}
\caption{(a) Plot of the absorptivity $\alpha (\theta, \omega)$ (red line) and the reflectivity $|r (\theta, \omega)|^2$ (blue line) as a function of the normalized frequency $\frac{\omega}{\omega_0}$. The
spectra are obtained for $Q=g=15$, i.e. for $\Gamma(\theta, \omega_0)=\omega_0$. 
(b) Plot of the normalized frequencies $\frac{\omega_\pm}{\omega_0}$ at half the maximum of the plasmon reflectivity $|r (\theta, \omega)|^2$ as a function of the ratio $\frac{\Gamma(\theta,\omega_0)}{\omega_0}$. 
In both panels, the full line represents the result of our complete model while the dashed line show the RWA calculation.}
\label{fig_ultra2}
\end{figure}

Figure~\ref{fig_ultra2}a presents absorptivity (red line) and reflectivity (blue line) spectra calculated in the rotating-wave approximation (dashed lines) and including anti-resonant terms of the light-matter interaction 
(continuous lines) in the ultra-strong coupling regime ($\Gamma(\theta, \omega_0)=\omega_0$ and $g=15$). The RWA leads to an overestimation of both absorptivity and reflectivity for $\omega > \omega_0$ and to an underestimation of 
these quantities at low frequencies. However, the most important difference between the RWA results and the full model concerns the high frequency limit of the optical coefficients. Indeed, in the RWA we find:
\begin{eqnarray*}
|r (\theta, \omega)|^2 &\underset{\omega \to \infty}{\xrightarrow{\hspace{1cm}}}& \frac{\Gamma (\theta,\omega_0)^2/4\omega_0^2}{1+\Gamma (\theta,\omega_0)^2/4\omega_0^2} \\
\end{eqnarray*}
Therefore very far from the resonance, the reflectivity tends to a constant value, which becomes significant when $\Gamma (\theta,\omega_0)  \sim \omega_0$. This result is obviously unphysical, as the two-dimensional electron gas 
should be fully transparent at high frequencies. Including anti-resonant terms, reflectivity and absorptivity both tend to zero for large values of $\omega$, so that high frequency radiation is entirely transmitted through the QW. 

This strong deviation from the rotating-wave approximation is the free-space analogue of that observed in the ultra-strong coupling regime between a matter excitation and a cavity mode~\cite{ciutiPRB2005ultrastrong}. In order to clarify the transition from the 
strong to the ultra-strong coupling regime between a multisubband plasmon and the free space radiation we plot in fig.~\ref{fig_ultra2}b the normalized frequencies at half the maximum $\omega_\pm$ of the reflectivity spectrum 
as a function of the ratio $\Gamma (\theta, \omega_0)/\omega_0$, with a comparison between the RWA (dashed line) and the full calculation results (full line). 
Above $\Gamma (\theta, \omega_0) = 0.3 \ \omega_0$, the frequencies at half the maximum calculated with the full model deviate from those obtained in the RWA. As a consequence $\Gamma (\theta, \omega_0) \approx 0.3 \, \omega_0$ 
is considered as the threshold for ultra-strong coupling. This is very similar to what is observed in the case of ultra-strong coupling with a microcavity, where anti-resonant terms are responsible for the non-linearity of 
the polariton energies. The upper branch in figure~\ref{fig_ultra2}b is more affected by anti-resonant terms than the lower one, as it follows the unphysical high-frequency 
behavior of the reflectivity. Finally, note that the condition $\Gamma (\theta, \omega_0) \approx \omega_0$ is experimentally accessible, as shown in ref.~\onlinecite{huppert_ACSph2015_Radiatively_broadened_incandescence}. 

\begin{figure}
\centering  
\includegraphics[scale=0.9]{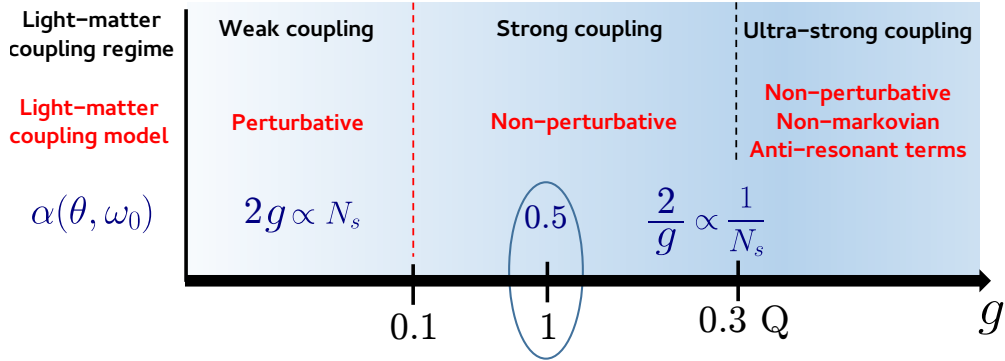}
\caption{Schematic representation of the different light-matter coupling regimes
identified in the text, as a function of the decay ratio $g$. The figure
also summarizes the main features that have to be accounted for in the 
quantum model to describe each regime, as well as the remarkable values
for the absorptivity at resonance $\alpha(\theta, \omega_0)$.}
\label{fig_recap}
\end{figure}

The progression between the three different light -- matter coupling regimes as a function of $g$ is represented in figure \ref{fig_recap}, which also summarizes the theoretical approaches describing each 
regime, together with the limits and remarkable values obtained for the absorptivity at resonance.

\section{Multisubband plasmon incandescence}
\label{sec_incandescence}
In the previous section we have investigated the linear optical properties of MSPs and identified the signatures of the ultra-strong coupling regime between a radiatively broadened material resonance and the free space radiation. 
Indications of ultra-strong coupling can also be searched in the properties of the radiation emitted under an electrical excitation. Our full quantum model is particularly suited to this aim. In particular it allows calculating the  
MSP incandescent emission arising when the photon and the electronic baths have different temperatures. In further work, the input-output formalism can be applied to study quantum properties of the emitted radiation in the presence of 
a modulation of the light-matter coupling, analogously to what has been done for intersubband polaritons, where such modulation has been predicted to give rise to dynamical Casimir 
effect~\cite{deliberatoPRL2007dynamic_casimir, auer_PRB2012_ultrastrong_casimir} and to the generation of squeezed radiation~\cite{fedortchenko2016_squeezing}. 
 
The incandescent emission process is described as a plasmon-mediated exchange between the electronic reservoir at temperature $T_e$ and the photon reservoir at temperature $T_{ph}$. 
Both baths are assumed to be in an incoherent thermal input state characterized by: 
\begin{eqnarray}
\langle b_{\mathbf k, \Omega}^{in \dag} b_{ \mathbf k', \Omega'}^{in} \rangle &=& n_B( \Omega, T_{el}) \  \delta_{\mathbf k}^{\mathbf k'} \delta (\Omega-\Omega')\label{eq_therm_b}\\
\langle a_{\mathbf k, s, \Omega}^{in \dag} a_{ \mathbf k', s', \Omega'}^{in} \rangle &=& n_B( \Omega, T_{ph}) \ \delta_{\mathbf k}^{\mathbf k'} \delta_s^{s'} \delta (\Omega-\Omega') \label{eq_therm_a}
\end{eqnarray}
with $n_B$ the Bose-Einstein occupancy: 
$$
n_B(\omega, T)= \frac{1}{e^{\frac{\hslash \omega}{k_B T}}-1}
$$
with $k_B$ the Boltzmann constant.
For such incoherent input, given that both $H_I^{ph}$ and $H_I^{el}$ conserve $\mathbf k$, the output field also verifies 
$\langle a_{\mathbf k, s, \Omega}^{out \dag} a_{ \mathbf k', s, \Omega'}^{out} \rangle \propto  \delta_{\mathbf k}^{\mathbf k'} $. 
Furthermore, it is naturally assumed that no correlation is present between the two baths in the input states, i.e. that products of $b^{in}$ and $a^{in}$ operators always have zero average value. 
Equation \eqref{eq_U} thus yields: 
\begin{eqnarray}
\langle a_{s, \Omega}^{out \dag} a_{s, \Omega'}^{out} \rangle &=&  n_B( \Omega, T_{ph}) \delta (\Omega-\Omega')\  \left[ |U_1^1(\Omega)|^2+|U_1^2(\Omega)|^2 \right]  \nonumber \\
&&+ \ n_B( \Omega, T_{el}) \delta (\Omega-\Omega')\ |U_1^{3}(\Omega)|^2 \label{Kir_text0} \\
 &=&  n_B( \Omega, T_{ph}) \delta (\Omega-\Omega')\  \left[ 1- \alpha(\Omega) \right]  \nonumber \\
&&+ \ n_B( \Omega, T_{el}) \delta (\Omega-\Omega')\ \alpha(\Omega) \label{Kir_text}
\end{eqnarray}
The second line of equation \eqref{Kir_text} is a direct consequence of the unitarity of matrix $U$. 

\begin{figure}
\centering  
\includegraphics[scale=0.7]{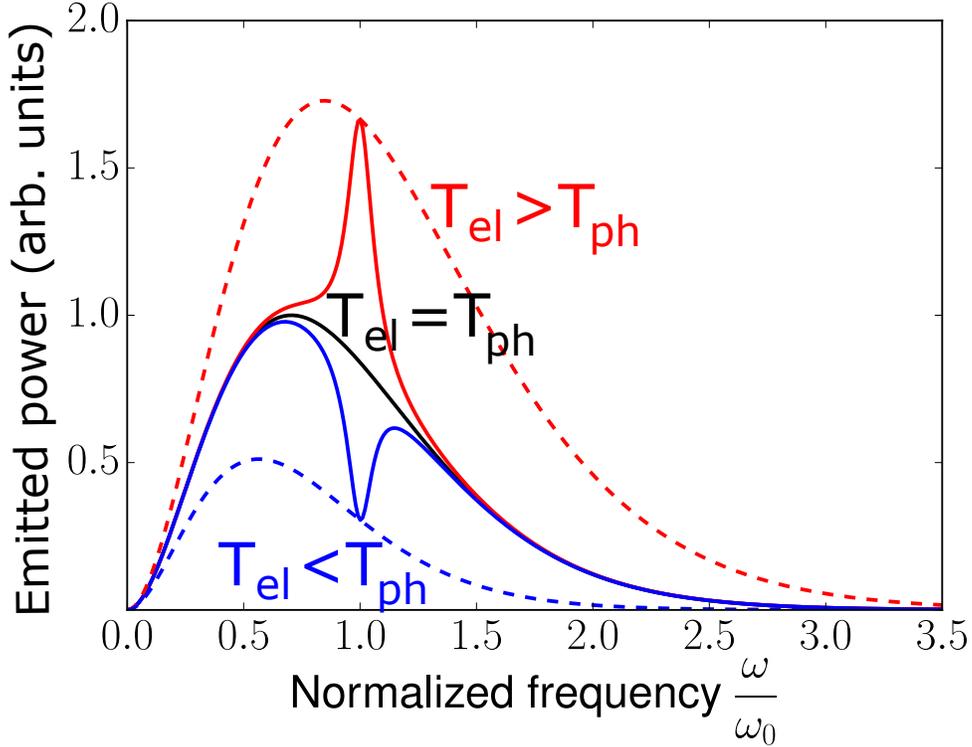}
\caption{Radiated power at the critical coupling angle for a MSP coupled with two baths at temperature $T_{el}$ and $T_{ph}$. The black line corresponds to Planck's law for black-body emission at thermal equilibrium $T_{el}=T_{ph}$. 
The continuous lines present the calculated radiation spectra when $T_{el}>T_{ph}$ (red line) or $T_{el}<T_{ph}$ (blue line). The dashed lines show Planck's emission at the temperature $T_{el}$ in the two cases.}
\label{fig_kirchhoff}
\end{figure}

For $T_{el}=T_{ph}$, the whole system is at thermal equilibrium, and absorption compensates emission exactly: the output photon number equals the input one (i.e. the thermal one). 
In the out-of-equilibrium case, equation \eqref{Kir_text} is equivalent to Kirchhoff's law of thermal emission: the first term on the right-hand side represents the number of input thermal photons 
remaining in the output (i.e. those that are not absorbed), whereas the second term corresponds to photon emission due to thermal fluctuations in the electronic bath. In agreement with Kirchhoff's law, the number of emitted 
photons is equal to the product of the absorptivity times the thermal occupancy at temperature $T_{el}$. 

This result is illustrated in figure~\ref{fig_kirchhoff}, presenting the emitted spectrum for a fixed temperature of the photon bath $T_{ph}$ and for three different electronic temperatures $T_{el}$. 
The spectra have been calculated for $g=\Gamma(\theta,\omega_0)/\gamma=1$, i.e. at the critical coupling angle, in the presence of a metallic mirror, such that the resonant absorptivity is equal to one. This situation has been experimentally studied 
in refs.~\onlinecite{laurentPRL2015_superradiance, huppert_ACSph2015_Radiatively_broadened_incandescence}. For $T_{ph}=T_{el}$ (black line) the entire system, including the plasmon and the two baths, is at thermal equilibrium 
and the number of emitted photons is simply given by the Bose distribution at the temperature of the system. For $T_{el}>T_{ph}$ (red line) the number of emitted photons deviates from the Bose distribution at temperature 
$T_{ph}$ and presents a peak at the MSP energy, where the absorptivity is unity and the number of emitted photons reaches the value $n_B(\omega_0, T_{el})$. For $T_{ph}>T_{el}$ the emitted spectrum presents a dip at the MSP energy: 
the plasmon transfers energy from the hot photonic bath to the cold electronic one.  

The model presented above has been successfully applied to reproduce plasmon emission experiments in which the two-dimensional electron gas is excited by an in-plane current in a device based on a field effect transistor geometry 
where the emitted light is extracted through a polished facet \cite{huppert_ACSph2015_Radiatively_broadened_incandescence}. In particular our model has accurately predicted the variations of the incandescent emission spectrum 
with the emission angle and the electronic density. These variations directly follow from the existence of the three regimes discussed in section \ref{sec_opt}. In the weak coupling regime, the emitted power increases linearly with $\Gamma(\theta, \omega_0) $ and thus with the electronic density~\cite{laurentPRL2015_superradiance}. When the critical coupling condition is met, the emissivity is maximum~\cite{huppert_ACSph2015_Radiatively_broadened_incandescence}. When 
$\gamma \ll \Gamma(\theta, \omega_0) \ll \omega_0, kT$, i.e. well into strong coupling but below the ultra-strong coupling limit, the emission spectrum is still Lorentzian, with a width $\Gamma(\theta, \omega_0) $ and 
an amplitude proportional to $1/\Gamma(\theta, \omega_0)$ (for the sake of simplicity, we assume $T_{ph}=0$). Therefore under these particular conditions, the integrated emitted power is independent on $\Gamma(\theta, \omega_0) $ and 
thus on both $N_s$ and $\theta$. Increasing the light-matter coupling strength $\Gamma(\theta, \omega_0) $ (for example by increasing the electronic density) only leads to a broadening and a flattening of the plasmon resonance, without affecting the total power. This peculiar behavior 
was also noted in a different superradiant system described in Ref. \onlinecite{PRB92_superradiant_thermal}, although it was not explicitly identified as a result of the strong coupling.

\section{Full diagonalization: properties of the light -- matter coupled states} 
\label{sec_GS}

In this section we present a different approach to study the ultra-strong coupling with free space radiation. We perform the full diagonalization of the coupled system Hamiltonian provided in section \ref{sec_H} and analyze the 
resulting properties of the system eigenstates, with particular attention to the ultra-strong coupling regime. As already discussed in section~\ref{sec_H}, the system consists of three bosonic reservoirs (two photonic and one electronic) with continuous energy spectra, coupled to a discrete bosonic state $P_{\mathbf k}$ (the multisubband plasmon). Therefore the eigenstates of the light -- matter coupled system can also be decomposed into three 
continua of bosonic excitations whose associated annihilation operators are denoted $\alpha_{\mathbf k, \Omega}$, $\alpha'_{\mathbf k, \Omega}$ and $\beta_{\mathbf k, \Omega}$. These operators must obey the characteristic equations: 
\begin{eqnarray}
\left[ \alpha'_{\mathbf k, \Omega}, H\right] &= &\hslash \Omega \ \alpha'_{\mathbf k, \Omega} \label{eq_alpha_p_H}\\
\left[ \beta_{\mathbf k, \Omega}, H\right] &=& \hslash \Omega \ \beta_{\mathbf k, \Omega} \label{eq_beta_H}   \\
\left[ \alpha_{\mathbf k, \Omega}, H\right] &=& \hslash \Omega \ \alpha_{\mathbf k, \Omega} \label{eq_alpha_H} 
\end{eqnarray}
Given the quadratic form of the Hamiltonian of section \ref{sec_H}, the bosonic excitations of the coupled system are linear combinations of the bare excitation operators: 
$P_{\mathbf k}$, $P_{-\mathbf k}^\dag$, $a_{\mathbf k, s, \Omega}$, $a_{-\mathbf k, s, \Omega}^\dag$, $b_{\mathbf k, \Omega}$ and $b_{-\mathbf k,  \Omega}^\dag$. 

We first show that it is possible to redefine the basis operators such that only one continuum couples with the MSP. For this we note that the photonic modes defined as 
$\alpha'_{\mathbf k, \Omega}= \frac{1}{\sqrt 2} \left[a_{\mathbf k, u, \Omega}-a_{\mathbf k, d, \Omega} \right]$ do not couple with $P_{\mathbf k}$. Therefore  
they follow equation \eqref{eq_alpha_p_H} and correspond to eigenstates of the coupled system. In a similar way, we identify the superposition 
$\beta_{\mathbf k, \Omega}$ of electronic and photonic modes that do not couple with the MSP:
\begin{equation}
 \beta_{\mathbf k, \Omega} = \frac{\sqrt{2\pi} K_{\mathbf k, \Omega}}{\sqrt{\gamma+\Gamma_{\mathbf k}(\Omega)}}\  a_{\mathbf k, \Omega} 
 -\frac{\sqrt{4\pi} W_{\mathbf k, \Omega}}{\sqrt{\gamma+\Gamma_{\mathbf k}(\Omega)}} \  b_{\mathbf k, \Omega} 
\end{equation}
In this equation, $a_{\mathbf k, \Omega}= \frac{1}{\sqrt 2} \left[a_{\mathbf k, u, \Omega}+a_{\mathbf k, d, \Omega} \right]$ is the continuum of photonic modes that is coupled to the plasmon, while $\Gamma_{\mathbf k}(\Omega)=4\pi  W_{\mathbf k, \Omega}^2 $ is the real part of the damping function $\tilde \Gamma_{\mathbf k}^{ph}(\Omega)$. The eigenmodes $\alpha'_{\mathbf k, \Omega}$ and 
$\beta_{\mathbf k, \Omega}$ are only defined above the light cone, i.e. for $\Omega>\frac{ck}{\sqrt{\epsilon_s}}$, while, below this limit, only the electronic excitations $b_{\mathbf k, \Omega}$ are present and 
coupled to the discrete plasmon mode. 

Having identified the uncoupled eigenstates $\alpha'$ and $\beta$, we show in appendix \ref{sec_app_diago} that the third continuum of eigenstates, the only one that is coupled with the plasmons, can be written as:
\begin{eqnarray}
 \alpha_{\mathbf k, \Omega} = &&f_{\mathbf k}(\Omega) \ \left[ P_{\mathbf k } + \frac{\Omega-\omega_0}{\Omega+\omega_0} \ P_{-\mathbf k }^\dag \right] \nonumber \\
 &+& \int d\Omega'\   g_{\mathbf k}(\Omega, \Omega') \bigg\{\  b_{\mathbf k, \Omega' } + \frac{\Omega'-\Omega}{\Omega'+\Omega} \ b_{-\mathbf k, \Omega' }^\dag  
 +                \      \frac{\sqrt{2}W_{\mathbf k,\Omega'}}{K_{\mathbf k,\Omega'}} \left[ \ a_{\mathbf k, \Omega' } + \frac{\Omega'-\Omega}{\Omega'+\Omega} \ a_{-\mathbf k, \Omega' }^\dag   \right]    \bigg\} \nonumber \\ \label{eq_alpha_def}
\end{eqnarray}
In this formula, function $g_{\mathbf k}$ is related to $f_{\mathbf k}$ by the following expression:
\begin{eqnarray}
g_{\mathbf k} (\Omega, \Omega') &=& -2i \frac{\omega_0 K_{\mathbf k,\Omega'}}{\Omega+\omega_0}  \left[\frac{1}{\Omega-\Omega'} +\pi \ z_{\mathbf k}(\Omega) \  \delta(\Omega-\Omega') \right]\  f_{\mathbf k}(\Omega)\label{eq_dirac_g} 
\end{eqnarray}
The function $z_{\mathbf k}(\Omega)$ depends on the frequency shift $G_{\mathbf k}(\Omega)=\text{Im}\left[ \tilde \Gamma_{\mathbf k}^{ph}(\Omega)-\tilde \Gamma_{\mathbf k}^{ph}(-\Omega)^*\right]$ 
(or Lamb shift, see section \ref{sec_QL}), according to:
\begin{equation}
 \Omega^2-\omega_0^2 -\omega_0 \ G_{\mathbf k}(\Omega) =\omega_0 \left[ \Gamma_{\mathbf k}(\Omega)+ \gamma \right] z_{\mathbf k}(\Omega)  \label{eq_z}
\end{equation}

In order to study the properties of the new eigenstates, described by the operators $\alpha_{\mathbf k, \Omega}$, we use the bosonic normalization conditions to derive the plasmon amplitude $f_{\mathbf k}$ 
(see appendix \ref{sec_app_diago}): 
\begin{equation}
 |f_{\mathbf k}(\Omega)|^2 = \frac{1}{2\pi} \frac{\gamma+\Gamma_{\mathbf k}(\Omega)}
 {\left[\Omega-\omega_0-\frac{\omega_0}{\omega_0+\Omega}\ G_{\mathbf k}(\Omega)\right]^2+ \frac{4\omega_0^2}{(\omega_0+\Omega)^2} \left[\frac{\gamma}{2} +\frac{\Gamma_{\mathbf k}(\Omega)}{2} \right]^2}  \label{eq_norm_f}
\end{equation}
As it was already mentioned in section \ref{sec_damp}, in the radiative region, i.e. for $\Omega>\frac{ck}{\sqrt{\epsilon_s}}$, the frequency shift $G_{\mathbf k}$ is zero, therefore the denominator in equation \eqref{eq_norm_f} 
is identical to the one appearing in the absorptivity, reflectivity and transmissivity formulae provided in section \ref{sec_opt}. 

\begin{figure}
\centering  
\includegraphics[scale=1.5]{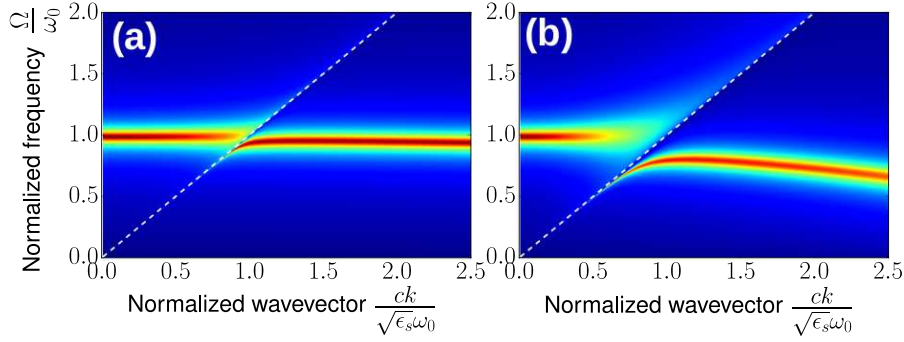}
\caption{Plot of the plasmon weight $|f(\Omega)|^2 \left[1-\frac{(\Omega-\omega_0)^2}{(\Omega+\omega_0)^2} \right]^2$ of the coupled system eigenstates $\alpha_{\mathbf k, \Omega}$, as a function of the normalized wavevector 
$\frac{ck}{\sqrt{\epsilon_s}\omega_0}$ and of the normalized frequency $\frac{\Omega}{\omega_0}$, for $\gamma=\frac{\omega_0}{15}$ and for $\Gamma_0=\frac{\omega_0}{30}$ (a) and $\Gamma_0=\frac{\omega_0}{6}$ (b). 
The dashed white line indicates the light cone: $\Omega=\frac{ck}{\sqrt{\epsilon_s}}$.}
\label{fig_dispersions}
\end{figure}

Figure~\ref{fig_dispersions} presents the plasmon weight (or plasmon Hopfield coefficient) of the light -- matter coupled operators $\alpha_{{\mathbf k}, \Omega}$, calculated as 
$|f_{\mathbf k}(\Omega)|^2 \left[1-\frac{(\Omega-\omega_0)^2}{(\Omega+\omega_0)^2} \right]$. In the figure the plasmon Hopfield coefficients are plotted in color scale as a function of the photon frequency and momentum 
(in normalized units) for coupling parameters typical of highly doped quantum well structures. In both panels, obtained for two different values of the ratio $\Gamma_0/\omega_0$, one can observe two modes: a radiative mode above the 
light-line (indicated by a dashed line in fig.~\ref{fig_dispersions}) and a non-radiative mode below the light line. 

The multisubband plasmon radiative mode broadens in the photon continuum when increasing $k$, as a result of the increase of the coupling constant $W_{\mathbf k, \Omega}$ between the plasmon and the photonic reservoir. This coupling constant presents a divergence at the light cone, i.e. for $\Omega=\frac{ck}{\sqrt{\epsilon_s}}$, which causes the system to enter the 
ultra-strong coupling regime, when $\Gamma_{\mathbf k}$ becomes a significant fraction of $\omega_0$. The broadening effect is stronger in fig. \ref{fig_dispersions}b, obtained for a higher ratio $\Gamma_0/\omega_0$: 
ultra-strong coupling is thus achieved on a larger domain. 

Note that the properties of the radiative mode perfectly reflect the characteristics of the optical spectra calculated within the input-output formalism and their dependence on the incidence angle. However, the present approach 
also provides information on the quantum eigenstates below the light cone, the non-radiative modes. Indeed all the relations derived above remain valid in that region except that 
the continuum of photon modes $a_{\mathbf k, \Omega}$ disappears, and therefore $\Gamma_{\mathbf k}(\Omega)=0$, while the energy shift $G_{\mathbf k}(\Omega)$ is now non-zero. In that case, equation \eqref{eq_norm_f} describes a Lorentzian resonance, 
with non-radiative linewidth $\gamma$, centered at the shifted energy $\omega_0+\frac{\omega_0}{\omega_0+\Omega}\ G_{\mathbf k}(\Omega)$. This resonance corresponds to a localized mode, also called 
\textit{Epsilon Near Zero mode}, and it has been described using a semiclassical approach in refs.~\onlinecite{alpeggianiPRB2014semiclassical_plasmons, Marquier_PRB2015_ENZ}. 
The two branches of the light -- matter coupled states are also very similar to those observed in $z$-polarized excitons in quantum wells~\cite{tassone}.

The annihilation operator describing the coupled state $\alpha_{\mathbf k, \Omega}$ is not only a combination of the annihilation operators of the uncoupled bosons, but also of the creation operators 
$P_{-\mathbf k}^\dag$, $a_{-\mathbf k, \Omega}^\dag$ and $b_{-\mathbf k, \Omega}^\dag$. This is a signature of the antiresonant coupling terms of the 
Hamiltonian. Due to the frequency ratios appearing in front of the creation operators in equation \eqref{eq_alpha_def}, these terms only become important when the broadening is a significant fraction of $\omega_0$, i.e. 
in the ultra-strong coupling regime. These creation operators are determinant for the investigation of the peculiar quantum properties of the ultra-strong coupling regime. Indeed, due to these terms, the ground state of 
ultra-strongly coupled systems contains a non-vanishing number of plasmons and photons. They are virtual excitations 
that can only be released if the light--matter coupling is modulated in time. This effect, called dynamical Casimir effect, was previously studied in the case of ultra--strong coupling between a material excitation and 
the discrete mode of an optical microcavity. In that case it was demonstrated that time-modulation of the Rabi frequency leads to the emission of an optical radiation outside the microcavity 
\cite{deliberatoPRL2007dynamic_casimir, auer_PRB2012_ultrastrong_casimir}. More recently, it was also shown that dynamical modulation of the Rabi frequency can generate squeezed light in the ultra--strong coupling 
regime\cite{fedortchenko2016_squeezing}. Our work shows that similar quantum effects could be observed in the case of ultra-strong coupling with free space radiation.

\section{Conclusion}
\label{sec:conclusion}
We demonstrated that ultra -- strong light -- matter coupling regime can exist in the absence of photonic confinement and be observed in systems in which the radiative broadening exceeds the non-radiative broadening and 
is comparable with the matter excitation energy. A system of choice for the observation of this regime is a dense two-dimensional electron gas, where collective electronic excitations with superradiant nature can be observed. 
We have developed an input-output model to study the optical properties of this system, including a microscopic description of the collective excitations. Our model, accounting consistently for radiative and non-radiative 
decay of collective excitations, clearly shows that anti -- resonant terms of the light -- matter interaction cannot be neglected and that RWA provides unphysical results at high incidence angles. 
We have also applied our formalism to the case of an electronic input, showing that Kirchhoff's law of thermal emission is a consequence of our general assumptions. 

The diagonalization of the complete light -- matter interaction Hamiltonian provided us with the mixed eigenstates of the coupled system, including a radiative mode and a localized epsilon near zero mode. 
We demonstrated that antiresonant terms of the interaction play an important role in the properties of the mixed states. In particular, due to these terms, the ground state of the ultra-strongly coupled system contains virtual 
plasmons and photons.
This work opens exciting perspectives, as it enables achieving ultra-strong coupling in systems free from any light confinement, and therefore it provides new ways to study the fascinating phenomena associated with this regime, 
such as dynamical Casimir effect\cite{deliberatoPRL2007dynamic_casimir, auer_PRB2012_ultrastrong_casimir}, ultraefficient light 
emission\cite{ciuti_PRA2006_inpout_output, deliberatoPRB2008_electroluminescence_polariton, deliberatoPRB2009_polariton_tunneling}, generation of squeezed light\cite{fedortchenko2016_squeezing} or enhanced charge 
transport\cite{feistPRL2015_exciton_conductance_USC, schachenmayerPRL2015_exciton_transport_cavity, orgiu}.

\acknowledgements
We acknowledge financial support from ERC (grant
ADEQUATE), Labex SEAM, and Agence
Nationale de la Recherche (grant ANR-14-CE26-0023-01). We
thank F. Alpeggiani and L. C. Andreani for fruitful discussions on the semi-classical theory of multisubband plasmons and S. Fedortchenko, P. Milman, T. Coudreau, A. Keller and C. Ciuti for several interesting discussions on the input-output model.

\newpage

\appendix
\begin{widetext}
\section{Superradiant light-matter coupling}
\label{sec_app_lamb}
\subsection{Real part of the damping function and emission rate}
To compute the spontaneous emission rate of the superradiant MSP, we start from the expression of light-matter coupling in the dipole representation of the Coulomb gauge\cite{todorovPRB2012plasmon_dipole_gauge}, considering that the MSP polarization is oriented 
along $z$: 
\begin{eqnarray}
H_I^{ph}&=& -\int d^3 \mathbf r \ \frac{D_z \cdot P_z}{\epsilon_0 \epsilon_s}  \label{eq_DdotP}
\end{eqnarray}
Using the standard quantization procedure for the electromagnetic field in a rectangular box with periodic boundary conditions, characterized by in-plane area $S$ and length $L_z$ along the growth axis, 
the displacement vector $D_z (\mathbf r)$ is rewritten as: 
\begin{eqnarray}
D_z(z, \mathbf r_{||})= i \sum_{q, \mathbf k} \sqrt{\frac{\hslash \epsilon_0 c^2 k^2 }{2SL_z \omega_{q,\mathbf k}}}  \   
e^{i\mathbf k \cdot \mathbf r_{||}+qz} \left[\alpha_{q, \mathbf k}-\alpha_{-q,- \mathbf k}^\dag \right]
\end{eqnarray}
In this expression, $q$ and $\mathbf k$ are the $z$ and in-plane wavevector components of the quantized photon mode characterized by the eigenfrequency $\omega_{q,\mathbf k}=\frac{c}{\sqrt{\epsilon_s}} \sqrt{k^2+q^2}$ 
(see fig \ref{schema}). The associated creation and annihilation operators are denoted $\alpha_{q, \mathbf k}$ and $\alpha_{q, \mathbf k}^\dag$. Using equation \eqref{eq_polar_plasm}, this yields:
\begin{eqnarray}
H_I^{ph} &=& i \hslash \sum_{n,q,\mathbf k} \nu_{n,q,\mathbf k} \left[ \alpha_{q,\mathbf k}^\dag -  \alpha_{q,-\mathbf k} \right] 
\left[ P_{n,\mathbf k}+ P_{n,-\mathbf k}^\dag   \right]  \label{eq_int_q}
\end{eqnarray}
The coupling constants $\nu_{n,q,\mathbf k}$ are computed in the long-wavelength approximation, i.e. assuming that the QW thickness is negligible compared to the optical wavelength $2\pi/q$. This yields:
\begin{equation}
\nu_{n,q,\mathbf k}= \frac{ c|\mathbf k| \sqrt{S} |\int dz \ J_n(z)|}{\omega_n\sqrt{2\epsilon_0 \epsilon_s^2 \hslash \omega_{q,\mathbf k} L_z}}   \label{eq_nu}
\end{equation}
In order to derive the Hamiltonian \eqref{eq_interaction}, it is necessary to change from the wavevector basis $(q, \mathbf k)$ to the frequency basis $(\Omega, \mathbf k)$. To that end we introduce new photon operators 
\begin{eqnarray}
a_{ \mathbf k, u, \Omega}&=& \frac{1}{\sqrt{\rho_{\mathbf k} (\Omega)}} \sum_{q>0} \alpha_{q,\mathbf k} \  \delta (\Omega-\omega_{q, \mathbf k}) \label{eq_a_up} \\
a_{ \mathbf k, d, \Omega}&=& \frac{1}{\sqrt{\rho_{\mathbf k} (\Omega)}} \sum_{q<0} \alpha_{q,\mathbf k} \  \delta (\Omega-\omega_{q, \mathbf k}) \label{eq_a_down} 
\end{eqnarray}
The photonic density of states $\rho_{\mathbf k}$, at fixed $\mathbf k$ is given by:
\begin{eqnarray}
\rho_{\mathbf k}(\Omega)&=& \frac{L_z}{2\pi} \frac{\epsilon_s}{c^2} \frac{\Omega}{\sqrt{\frac{\epsilon_s \Omega^2}{c^2}- k^2}} \Theta \left(\frac{\epsilon_s \Omega^2}{c^2}- k^2 \right)  \label{rho}
\end{eqnarray}
With this definition, it can be easily checked that the operators $a_{ \mathbf k, s, \Omega}$ and $a_{ \mathbf k, s, \Omega}^\dag$ still follow bosonic commutation rules and that the full system Hamiltonian takes the form given in 
section \ref{sec_H}, with the following definition for the renormalized coupling coefficients: 
\begin{equation}
W_{n, \mathbf k,  \Omega}= \sqrt{\rho_{\mathbf k} (\Omega)} \  \nu_{n,q,\mathbf k}  \label{eq_W}
\end{equation}

In the case of highly doped QWs, one superradiant MSP mode $n=0$ concentrates most of the oscillator strength. Its radiative decay rate is expressed as:
\begin{eqnarray}
\text{Re}  \left[ \tilde \Gamma^{ph}_{\mathbf k } (\omega) \right] &=& 4\pi \ W_{0, \mathbf k, \omega}^2   \label{app_realGam} \\
&=&  \Gamma_0 \frac{ ck^2}{\sqrt{\epsilon_s} \omega_0 \sqrt{\frac{\epsilon_s \omega^2}{c^2}- k^2}} \Theta \left(\frac{\epsilon_s \omega^2}{c^2}- k^2 \right) \label{app_realGammafinal}
\end{eqnarray}
with
$$
\Gamma_0= \frac{S}{\hslash \epsilon_0 \sqrt{\epsilon_s}} \frac{|\int dz \ J_0(z)|^2}{c \omega_0}
$$
As shown in section \ref{sec_MSP}, $\Gamma_0$ is approximately proportional to the surface electronic density $N_s$\cite{laurentPRL2015_superradiance}.

\subsection{Imaginary part of the damping function and Lamb shift}

The imaginary part of the radiative damping function is obtained from equation \eqref{gam_ph_t} and verifies the following Kramers-Kronig relation: 
\begin{eqnarray}
\text{Im}  \left[ \tilde \Gamma^{ph}_{\mathbf k } (\omega) \right]  &=&  
\frac{1}{\pi} P \int_{-\infty}^{\infty} \ d\omega' \ \frac{\text{Re}  \left[ \tilde \Gamma^{ph}_{\mathbf k } (\omega') \right]}{\omega-\omega'} \label{imagGam}
\end{eqnarray}
This term corresponds to a Lamb shift of the plasmon frequency due to emission and absorption of virtual photons. This shift can be evaluated analytically combining equations \eqref{app_realGammafinal} and \eqref{imagGam}:
\begin{eqnarray*}
\text{Im}  \left[ \tilde \Gamma^{ph}_{\mathbf k }  (\omega) \right]  &=&\frac{ \Gamma_0 c k^2}{ \pi \omega_0 \sqrt{\epsilon_s}}
P \int_{\frac{ck}{\sqrt{\epsilon_s}}}^{\infty} \ d\omega' \ \frac{1}{(\omega-\omega')\sqrt{\frac{\epsilon_s \omega'^2}{c^2}- k^2}} \\
  &=&\frac{ 2\Gamma_0 c}{ \pi \omega_0 \sqrt{\epsilon_s}} \frac{k^2}{\sqrt{ \frac{\epsilon_s \omega^2}{c^2}-k^2}} 
 \log\left[\sqrt{\frac{\omega+\frac{ck}{\sqrt{\epsilon_s}} }{2 \frac{ck}{\sqrt{\epsilon_s}} }}+\sqrt{\frac{\omega-\frac{ck}{\sqrt{\epsilon_s}} }{2 \frac{ck}{\sqrt{\epsilon_s}} }} \right] \  
 \text{if} \  \omega>\frac{ck}{\sqrt{\epsilon_s}}  \\
&=& - \frac{ \Gamma_0 c }{ \pi \omega_0 \sqrt{\epsilon_s}} \frac{k^2}{\sqrt{ k^2-\frac{\epsilon_s \omega^2}{c^2}}} 
 \left[\frac{\pi}{2}+ \sin^{-1} \left(\frac{\sqrt{\epsilon_s}\omega}{ck} \right) \right] \  \text{if} \  -\frac{ck}{\sqrt{\epsilon_s}} < \omega<\frac{ck}{\sqrt{\epsilon_s}}  \\
  &=&\frac{ 2\Gamma_0 c}{ \pi \omega_0 \sqrt{\epsilon_s}} \frac{k^2}{\sqrt{ \frac{\epsilon_s \omega^2}{c^2}-k^2}} 
 \log\left[\sqrt{\frac{|\omega|+\frac{ck}{\sqrt{\epsilon_s}} }{2 \frac{ck}{\sqrt{\epsilon_s}} }}-\sqrt{\frac{|\omega|-\frac{ck}{\sqrt{\epsilon_s}} }{2 \frac{ck}{\sqrt{\epsilon_s}} }} \right] \  
 \text{if} \  \omega<-\frac{ck}{\sqrt{\epsilon_s}}  \\
\end{eqnarray*}
In the radiative region, i.e. when $|\omega|>\frac{ck}{\sqrt{\epsilon_s}}$, the imaginary part of  $\tilde \Gamma^{ph}_{\mathbf k }$ is much lower than its real part and it is an odd function of $\omega$. 
Therefore the corresponding frequency shift $\text{Im}  \left[ \tilde \Gamma^{ph}_{\mathbf k }  (\omega) -\tilde \Gamma^{ph*}_{\mathbf k }  (-\omega) \right]$ cancels. 

In the non-radiative region, i.e. when $|\omega|<\frac{ck}{\sqrt{\epsilon_s}}$, the frequency shift is non-zero and it determines the energy of the localized mode (or epsilon near zero mode). 
It is given by the following relation:
\begin{eqnarray}
\text{Im}  \left[ \tilde \Gamma^{ph}_{\mathbf k } (\omega) -\tilde \Gamma^{ph*}_{\mathbf k } (-\omega)\right] 
&=&  -\Gamma_0 \frac{ ck^2}{\sqrt{\epsilon_s} \omega_0 \sqrt{ k^2-\frac{\epsilon_s \omega^2}{c^2}}} \Theta \left(k^2-\frac{\epsilon_s \omega^2}{c^2} \right) \label{eq_app_imagGammafinal}
\end{eqnarray}
For the analysis of optical properties of highly doped quantum wells provided in sections \ref{sec_model} to \ref{sec_incandescence}, the only relevant values of $\omega$ are in the radiative region, therefore the frequency shift is 
discarded. However, equation \eqref{eq_app_imagGammafinal} relies on the long-wavelength approximation. This approximation breaks down in thick heterostructures, 
in which the Lamb shift might become observable, and is expected to be enhanced due to the superradiant nature of multisubband plasmons~\cite{rohlsberger, scully}. 

\section{Diagonalization of the ultra-strong coupling Hamiltonian}
\label{sec_app_diago}

In this appendix we provide the details of the diagonalization procedure used in section \ref{sec_GS}. 
Once the uncoupled continua $\alpha'_{\mathbf k, \Omega}$ and $\beta_{\mathbf k, \Omega}$ have been identified, we describe the annihilation operators $\alpha_{\mathbf k, \Omega}$ of the mixed eigenstates with the general expression: 
\begin{eqnarray}
 \alpha_{\mathbf k, \Omega} = &&f_{\mathbf k}(\Omega) \ P_{\mathbf k } + \tilde f_{\mathbf k}(\Omega) \ P_{-\mathbf k }^\dag \nonumber \\
 &+& \int d\Omega'\  \bigg[ g_{\mathbf k}(\Omega, \Omega')\  b_{\mathbf k, \Omega' } + \tilde g_{\mathbf k}(\Omega, \Omega') \ b_{-\mathbf k, \Omega' }^\dag 
 +                     h_{\mathbf k}(\Omega, \Omega') \ a_{\mathbf k, \Omega' } + \tilde h_{\mathbf k}(\Omega, \Omega') \ a_{-\mathbf k, \Omega' }^\dag       \bigg] \nonumber \\ \label{eq_app_alpha_def}
\end{eqnarray}
We then introduce this definition into the eigenstate commutation relation:
\begin{equation}
 \left[ \alpha_{\mathbf k, \Omega}, H\right] = \hslash \Omega \ \alpha_{\mathbf k, \Omega}  \label{eq_app_alpha_om} 
\end{equation}
By equating the coefficients of the basis operators appearing on both sides of \eqref{eq_app_alpha_om}, we derive the following relations (the index $\mathbf k$ is omitted in the following when unnecessary): 
\begin{eqnarray}
&&\tilde f(\Omega)= f(\Omega) \times \frac{\Omega-\omega_0}{\Omega+\omega_0} \label{eq_tilde_f}\\
&&\tilde g(\Omega, \Omega')= g(\Omega, \Omega') \times \frac{\Omega'-\Omega}{\Omega'+\Omega} \label{eq_tilde_g}\\
&&\tilde h(\Omega, \Omega')= h(\Omega, \Omega') \times \frac{\Omega'-\Omega}{\Omega'+\Omega} \label{eq_tilde_h} 
\end{eqnarray}
\begin{eqnarray}
&& \big[\Omega-\Omega'\big] g (\Omega, \Omega')  =-i \frac{2\omega_0}{\Omega+\omega_0}  K_{\Omega'}  \   f(\Omega)\label{eq_f_g} \\ 
&&\big[\Omega-\Omega'\big]  h (\Omega, \Omega') =-i \frac{2\omega_0}{\Omega+\omega_0} \sqrt{2 } W_{\Omega'} \  f(\Omega)\label{eq_f_h} \\ 
&&\big[\Omega-\omega_0\big]  f (\Omega) =i \int d\Omega' \   \frac{2\Omega' }{\Omega+\Omega'} K_{\Omega'}\  g (\Omega, \Omega')+ \frac{2\Omega'}{\Omega+\Omega'} \sqrt{2} W_{\Omega'}\  h (\Omega, \Omega')   \label{eq_f_g_h}
\end{eqnarray}
To write these expressions, we assumed for simplicity that $K_{ \Omega}$ is real but the general results derived in this section are also true for complex values of $K_{\Omega}$. To solve this system of equations, we use Dirac's 
method\cite{fanoPR1961} and replace equations \eqref{eq_f_g} and \eqref{eq_f_h} with the following ansatz:
\begin{eqnarray}
g (\Omega, \Omega') &=&-2i \frac{\omega_0 K_{\Omega'}}{\Omega+\omega_0} \left[ \frac{1}{\Omega-\Omega'} + \pi \ z_{el}(\Omega) \  \delta(\Omega-\Omega') \right]\  f(\Omega)\label{eq_app_dirac_g} \\ 
h (\Omega, \Omega') &=&-2i \frac{\sqrt{2} \omega_0 W_{\Omega'}}{\Omega+\omega_0}\left[ \frac{1}{\Omega-\Omega'} + \pi \  z_{ph}(\Omega) \  \delta(\Omega-\Omega') \right]\  f(\Omega)\label{eq_app_dirac_h} 
\end{eqnarray}
The $\delta$-function terms are added to treat the equality case $\Omega=\Omega'$ (when $g$ or $h$ are integrated over $\Omega'$, only the principal part of the divergent fraction $[\Omega-\Omega' ]^{-1}$ should be considered). 
Furthermore, inserting equations \eqref{eq_app_dirac_g} and \eqref{eq_app_dirac_h} into the orthogonality condition $[\beta_{\Omega'}, \alpha_{\Omega}^\dag]=0$, we find that $z_{el}(\Omega)=z_{ph}(\Omega)\equiv z(\Omega)$. 
These expressions are then injected into \eqref{eq_f_g_h} and yield the following relation (from which the value of $z(\Omega)$ can be inferred): 
\begin{equation}
 \Omega^2-\omega_0^2 = \omega_0 \left[\gamma + \Gamma(\Omega)\right] z(\Omega) + \omega_0 G_{el}(\Omega) + \omega_0 G_{ph}(\Omega)   \label{eq_app_z_12}
\end{equation}
Where,
\begin{eqnarray}
 G_{el}(\Omega)&=& \int d\Omega' \  \frac{4\Omega'K_{\Omega'}^2 }{\Omega^2-\Omega'^2} =
 \frac{1}{\pi}\left\{\text{Im}  \left[ \tilde \Gamma^{el} (\Omega) \right]+\text{Im}  \left[ \tilde \Gamma^{el} (-\Omega) \right] \right\} \\
 G_{ph}(\Omega)&=& \int d\Omega' \  \frac{8\Omega'W_{\Omega'}^2 }{\Omega^2-\Omega'^2} =
 \frac{1}{\pi} \left\{\text{Im}  \left[ \tilde \Gamma^{ph} (\Omega) \right]+\text{Im}  \left[ \tilde \Gamma^{ph} (-\Omega) \right] \right\} 
\end{eqnarray}
Following the analysis carried out in appendix \ref{sec_app_lamb}, we neglect the frequency shift $G_{el}$, while $G_{ph}$ is given by equation \eqref{eq_app_imagGammafinal}. We have so far demonstrated the relations 
\eqref{eq_alpha_def}, \eqref{eq_dirac_g} and \eqref{eq_z}. The only remaining step to characterize completely the eigenstates $\alpha_{\mathbf k, \Omega}$, is to determine the plasmon amplitude $f(\Omega)$. 
This is done using equations \eqref{eq_tilde_f}, \eqref{eq_tilde_g} and \eqref{eq_tilde_h}, together with the ansatz \eqref{eq_app_dirac_g} and \eqref{eq_app_dirac_h} in order to rewrite $\alpha_{ \Omega}$ 
as a function of the only remaining unknown parameter $f(\Omega)$. We then apply the following commutation rule: 
\begin{eqnarray}
 \delta(\Omega_1-\Omega_2) &=& [\alpha_{\Omega_1}, \alpha_{\Omega_2}^\dag]\nonumber \\
&=& f(\Omega_1) f(\Omega_2)^* \ \frac{2\omega_0 (\Omega_1+\Omega_2)}{(\Omega_1+\omega_0)(\Omega_2+\omega_0)} \times  \bigg\{ 1 + \int d\Omega \ \left[1+\frac{\Gamma(\Omega)}{\gamma} \right] 
\frac{4\Omega \omega_0 K_{\Omega}^2}{(\Omega_1+\Omega)(\Omega_2+\Omega)}  \nonumber \\
&\times& \left[ \frac{1}{\Omega_1-\Omega} + \pi z(\Omega_1) \ \delta(\Omega_1-\Omega) \right] \left[ \frac{1}{\Omega_2-\Omega} + \pi z(\Omega_2) \  \delta(\Omega_2-\Omega) \right] \bigg\}\label{eq_app_norm1}
\end{eqnarray}
In the integral over $\Omega$ of the equation above, the terms involving $\delta$-functions are easily computed. The last term is: 
\begin{eqnarray}
\frac{\omega_0}{\pi} \int 2\Omega d\Omega \ \left[\gamma+\Gamma(\Omega) \right] \frac{1}{\Omega_1^2-\Omega^2} \frac{1}{\Omega_2^2-\Omega^2} 
&=& \frac{\omega_0}{\pi} \int du \ \left[\gamma+\Gamma(\sqrt{u}) \right] \  \frac{1}{\Omega_1^2-u} \frac{1}{\Omega_2^2-u} \nonumber \\
&=& \omega_0\frac{G_{el}(\Omega_2)+G_{ph}(\Omega_2)-G_{el}(\Omega_1)-G_{ph}(\Omega_1)}{\Omega_1^2-\Omega_2^2} \nonumber \\
&+& \frac{\pi \omega_0}{2\Omega_1} \left[\gamma+\Gamma(\Omega_1) \right] \ \delta(\Omega_1-\Omega_2) \label{eq_app_int_u}
\end{eqnarray}
For this last equality, we have used the following relation on principal parts of divergent fractions\cite{fanoPR1961}:
\begin{equation}
 \frac{1}{\Omega_1^2-u} \frac{1}{\Omega_2^2-u} = \frac{1}{\Omega_1^2-\Omega_2^2} \left[\frac{1}{\Omega_2^2-u} - \frac{1}{\Omega_1^2-u} \right] + \pi^2 \ \delta(u- \Omega_1^2)  \ \delta(\Omega_1^2- \Omega_2^2) 
\end{equation}
Finally, in the right-hand side of equation \eqref{eq_app_norm1}, all terms cancel but the ones proportional to $\delta(\Omega_1-\Omega_2)$ (this is a consequence of equation \eqref{eq_app_z_12} applied at 
$\Omega=\Omega_1$ and $\Omega=\Omega_2$). We thus obtain the following equality for $|f(\Omega)|$, which is equivalent to \eqref{eq_norm_f}:
\begin{equation}
 |f(\Omega)|^2 \  \frac{2\pi \omega_0^2}{(\Omega+\omega_0)^2} \left[ \gamma+\Gamma(\Omega) \right] \left[1+z(\Omega)^2\right] =1 \label{eq_app_norm_f0}
\end{equation}
Note that all the expressions above, although derived above the light cone, remain valid for $\Omega< \frac{ck}{\sqrt{\epsilon_s}}$.

\end{widetext}

\bibliographystyle{apsrev4-1}
\bibliography{biblio}

\end{document}

%% file: fig_plasm4.eps_tex
\begingroup%
  \makeatletter%
  \providecommand\color[2][]{%
    \errmessage{(Inkscape) Color is used for the text in Inkscape, but the package 'color.sty' is not loaded}%
    \renewcommand\color[2][]{}%
  }%
  \providecommand\transparent[1]{%
    \errmessage{(Inkscape) Transparency is used (non-zero) for the text in Inkscape, but the package 'transparent.sty' is not loaded}%
    \renewcommand\transparent[1]{}%
  }%
  \providecommand\rotatebox[2]{#2}%
  \ifx\svgwidth\undefined%
    \setlength{\unitlength}{283.79211426bp}%
    \ifx\svgscale\undefined%
      \relax%
    \else%
      \setlength{\unitlength}{\unitlength * \real{\svgscale}}%
    \fi%
  \else%
    \setlength{\unitlength}{\svgwidth}%
  \fi%
  \global\let\svgwidth\undefined%
  \global\let\svgscale\undefined%
  \makeatother%
  \begin{picture}(1,0.56722859)%
    \put(0,0){\includegraphics[width=\unitlength]{fig_plasm4.eps}}%
    \put(0.12663148,0.55536125){\color[rgb]{0,0,0}\makebox(0,0)[lt]{\begin{minipage}{0.0185013\unitlength}\raggedright $z$\end{minipage}}}%
    \put(0.06976072,0.45186973){\color[rgb]{0,0,0}\makebox(0,0)[lt]{\begin{minipage}{0.02085622\unitlength}\raggedright $x$\end{minipage}}}%
    \put(0.27535711,0.45368114){\color[rgb]{0,0,0}\makebox(0,0)[lt]{\begin{minipage}{0.01989048\unitlength}\raggedright $y$\end{minipage}}}%
    \put(0.1939938,0.18148411){\color[rgb]{0,0,0}\makebox(0,0)[lt]{\begin{minipage}{0.06629064\unitlength}\raggedright $q$\end{minipage}}}%
    \put(0.26963071,0.09320095){\color[rgb]{0,0,0}\makebox(0,0)[lt]{\begin{minipage}{0.1518134\unitlength}\raggedright $\mathbf k$\end{minipage}}}%
    \put(0.24704607,0.19986125){\color[rgb]{0,0,0}\makebox(0,0)[lt]{\begin{minipage}{0.05094065\unitlength}\raggedright $\theta$\end{minipage}}}%
    \put(0.44520719,0.16879722){\color[rgb]{0,0,0}\makebox(0,0)[lb]{\smash{$a_{\mathbf k, u, \Omega}^{in}$}}}%
    \put(0.79663261,0.17111241){\color[rgb]{0,0,0}\makebox(0,0)[lb]{\smash{$a_{\mathbf k, d, \Omega}^{out}$}}}%
    \put(0.72866224,0.52993593){\color[rgb]{0,0,0}\makebox(0,0)[lb]{\smash{$a_{\mathbf k, u, \Omega}^{out}$}}}%
    \put(0.86382313,0.32022393){\color[rgb]{0,0,0}\makebox(0,0)[lb]{\smash{$P_{n,\mathbf k}$}}}%
    \put(0.49000712,0.42618163){\color[rgb]{0,0,0}\makebox(0,0)[lb]{\smash{$\lambda/2=\pi/k$}}}%
  \end{picture}%
\endgroup%

%% file: fig_system.eps_tex
\begingroup%
  \makeatletter%
  \providecommand\color[2][]{%
    \errmessage{(Inkscape) Color is used for the text in Inkscape, but the package 'color.sty' is not loaded}%
    \renewcommand\color[2][]{}%
  }%
  \providecommand\transparent[1]{%
    \errmessage{(Inkscape) Transparency is used (non-zero) for the text in Inkscape, but the package 'transparent.sty' is not loaded}%
    \renewcommand\transparent[1]{}%
  }%
  \providecommand\rotatebox[2]{#2}%
  \ifx\svgwidth\undefined%
    \setlength{\unitlength}{267.77448176bp}%
    \ifx\svgscale\undefined%
      \relax%
    \else%
      \setlength{\unitlength}{\unitlength * \real{\svgscale}}%
    \fi%
  \else%
    \setlength{\unitlength}{\svgwidth}%
  \fi%
  \global\let\svgwidth\undefined%
  \global\let\svgscale\undefined%
  \makeatother%
  \begin{picture}(1,0.52124521)%
    \put(0,0){\includegraphics[width=\unitlength]{fig_system.eps}}%
    \put(0.07405712,0.34833097){\color[rgb]{0,0,0}\makebox(0,0)[lt]{\begin{minipage}{0.50682163\unitlength}\raggedright $\Omega$\end{minipage}}}%
    \put(0.05563862,0.46869991){\color[rgb]{0,0,0}\makebox(0,0)[lt]{\begin{minipage}{0.34277468\unitlength}\raggedright \center{Electronic bath\\ $b_{n,\mathbf k, \Omega}, b_{n,\mathbf k, \Omega}^\dag$}\end{minipage}}}%
    \put(0.41421131,0.52443296){\color[rgb]{0,0,0}\makebox(0,0)[lt]{\begin{minipage}{0.23985065\unitlength}\raggedright \center{Multisubband \\ plasmons\\ $P_{n,\mathbf k}, P_{n,\mathbf k}^\dag$}\end{minipage}}}%
    \put(0.66463523,0.46564148){\color[rgb]{0,0,0}\makebox(0,0)[lt]{\begin{minipage}{0.35388517\unitlength}\raggedright \center{Photon bath\\ $a_{\mathbf k, s, \Omega}, a_{\mathbf k, s, \Omega}^\dag$}\end{minipage}}}%
    \put(0.58596548,0.1191354){\color[rgb]{0,0,0}\makebox(0,0)[lt]{\begin{minipage}{0.21814341\unitlength}\raggedright \center{Coupling}\end{minipage}}}%
    \put(0.28334441,0.11801421){\color[rgb]{0,0,0}\makebox(0,0)[lt]{\begin{minipage}{0.23081874\unitlength}\raggedright \center{Coupling }\end{minipage}}}%
    \put(0.28307402,0.07146033){\color[rgb]{0,0,0}\makebox(0,0)[lt]{\begin{minipage}{0.21585328\unitlength}\raggedright \center{$K_{n,\mathbf k, \Omega}$}\end{minipage}}}%
    \put(0.56988253,0.07071344){\color[rgb]{0,0,0}\makebox(0,0)[lt]{\begin{minipage}{0.23713976\unitlength}\raggedright \center{$W_{n,\mathbf k, \Omega}$}\end{minipage}}}%
    \put(0.05733164,0.26309805){\color[rgb]{0,0,0}\makebox(0,0)[lt]{\begin{minipage}{0.08706543\unitlength}\raggedright $\omega_n$\end{minipage}}}%
  \end{picture}%
\endgroup%